\documentclass[aps,prb,twocolumn,superscriptaddress]{revtex4-2}
\usepackage{amssymb}
\usepackage{amsmath}
\usepackage{mathrsfs}
\usepackage{mathtools}
\mathtoolsset{showonlyrefs}
\usepackage{bm}
\usepackage{physics}
\usepackage[svgnames]{xcolor}
\usepackage{float}
\usepackage{tikz}
\usepackage{graphicx}
\usepackage[
    setpagesize=false,
    bookmarks=false,
    colorlinks=true,
    allcolors=magenta
]{hyperref} 
\usepackage[titletoc]{appendix}

\begin{document}

\title{Qubit dynamics beyond Lindblad:\\ Non-Markovianity versus rotating wave approximation}

\author{Kiyoto Nakamura}
\email{kiyoto.nakamura@uni-ulm.de}
\author{Joachim Ankerhold}
\affiliation{Institute for Complex Quantum Systems and IQST, Ulm University, D-89069 Ulm, Germany}
\date{\today}

\begin{abstract}
\textcolor{black}{With increasing performance of actual qubit devices, even subtle effects in the interaction between qubits and environmental degrees of freedom become progressively relevant and experimentally visible. This applies particularly to the timescale separations that are at the basis of the  most commonly used numerical simulation platform for qubit operations, namely, the conventional Lindblad master equation (LE): the Markov approximation and the rotating wave approximation (RWA).  
In this contribution, we shed light on the questions (i) to which extent it is possible to monitor violations of either of these timescale separations experimentally and (ii) which of them is the most severe to provide highly accurate predictions within (approximate) numerical schemes in relevant parameter ranges. For this purpose, we compare three simulation methods for the reduced density matrix with progressively growing accuracy.
In particular, predictions for relaxation and decoherence of a qubit system in the presence of reservoirs with Ohmic and sub-Ohmic spectral densities are explored and, with the aid of proper protocols based on Ramsey experiments, the role of non-Markovianity and RWA are revealed. We discuss potential implications for future experiments and the design of approximate yet accurate numerical approaches.}
\end{abstract}

\maketitle

\section{Introduction}
One of the major obstacles to realize universal quantum computing is the omnipresence of decoherence. Impressive progress has been achieved in the last decade in terms of coherence times~\cite{PlaceNATCOMMUN2021,WangNPJQI2022} and gate fidelities~\cite{NegirneacPRL2021,SungPRX2021,KandalaPRL2021} but particularly for implementations on solid state platforms such as superconducting circuits and semiconducting devices, larger arrays of qubits still suffer from residual noise sources \cite{GoogleNATURE2019,GoogleNATURE2023,IBMNATURE2023}. However, the susceptibility of single and few qubit devices has approached a level where even minute details of environmental effects can be monitored, e.g.,\ cosmic radiation and radioactivity \cite{CardaniNATCOMMUN2021}. While the latter appear as rare events on intrinsic qubit timescales, subtle details of ubiquitous broadband noise, effective also on short and moderate timescales, have turned into the focus to improve circuit designs and protocols \cite{TuorilaPRR2019}.

A very powerful approach to include noise in a quantum dynamical setting is a description of qubit operations in terms of the Lindblad master equations (LE)~\cite{LindbladCMP1976}. However, these come with severe limitations such that numerical predictions may no longer match fidelities of the experimental performance~\cite{PapicARXIV2023}. 
Roughly speaking, for weak system--reservoir couplings, a prerequisite for qubit operations, the LE follows from the general framework of open quantum systems in terms of system+reservoir models by assuming two types of timescale separations~\cite{Breuer2002,Gardiner2010,Weiss2012}. The first appears in the interaction picture and implies a timescale separation between the dynamics of the reduced density operator and the decay of correlations in the reservoir. This leads effectively to a time-local equation of motion for the reduced density operator (Born--Markov approximation), the so-called Bloch--Redfield equation~\cite{RedfieldIJRD1957}, which in general is not of Lindblad form though. The LE appears through an additional time coarse graining on timescales $t\gg 1/\omega_q$ with qubit transition frequency $\omega_q$. This second time coarse graining is consistently performed in the eigenbasis of the system Hamiltonian and boils down to a rotating wave approximation (RWA), also known as secular approximation, where rapid oscillations of the system are neglected. Hence, conventional LEs not only come with limited ranges of applicability (sufficiently elevated temperatures, weak coupling) and, within this range, with limited predictive power in terms of accuracy.

While the Born--Markov approximation and RWA go hand in hand in the LE, one may wonder whether they lead to distinct phenomena that can be distinguished based on the improved sensing capabilities of advanced qubit devices. 
As an example, we mention 
a recent study~\cite{GulasciPRB2022}, 
where an experimental protocol is proposed to monitor non-Markovianity. 
\textcolor{black}{However, it remains still unclear which timescale separation, Born--Markov or RWA, is the most severe one in the parameter ranges, where qubits are operated. To which extent is it possible to observe violations of either of the timescale separations experimentally?  Which one should be avoided the most in theoretical simulations if this can be said at all?}

\textcolor{black}{Here, we provide an analysis which sheds light onto this issue by comparing predictions from three different simulation platforms which are chosen according to their progressively growing level of accuracy: The conventional LE involves the Born--Markov and the RW approximation, the so-called universal Lindblad equation (ULE)~\cite{NathanPRB2020} involves only the Born--Markov approximation, while the 
recently extended version of the Hierarchical Equation of Motion (FP-HEOM) produces exact benchmarks (full non-Markovianity, no RWA) down to temperature zero \cite{heomHigh,XuPRL2022}. The goal of this contribution is thus {\em not} to provide a comprehensive comparison of various approximate simulations schemes for open quantum systems with exact data but rather to identify the relevance of the respective timescale separations for realistic qubit descriptions. Our findings may then allow on the experimental side to develop protocols to monitor deviations from Markovian resp.\ RWA qubit dynamics (an interesting issue on its own) and on the theoretical side to improve approximate but computationally less expensive schemes to reliably predict multiqubit operations.}

This paper is organized as follows. In Sec.~\ref{sec:deriveLE}, we review the derivation of the conventional Lindblad equation and universal Lindblad equation. In the following, we simply refer to the conventional Lindblad equation as the Lindblad equation. We derive the HEOM in Sec.~\ref{sec:deriveHEOM}. In Sec.~\ref{sec:results}, we demonstrate the numerical results obtained with these three methods, and study the differences caused by the Born--Markov approximations and RWA.
We also propose a new experimental protocol for the detection of the non-Markovianity there.
As a model, a two-level system is considered.
Section~\ref{sec:conclusion} is devoted to the concluding remarks.

\section{Derivation of the Lindblad equations} \label{sec:deriveLE}
To derive the Lindblad equation (LE) and universal Lindblad equation (ULE), we start from the Caldeira--Leggett Hamiltonian that is in the following form:
\begin{align}
    \hat{H}_\mathrm{tot} & = \hat{H}_S +
    \sum_j \left(\frac{\hat{p}_j^2}{2m_j} + 
    \frac{1}{2} m_j \omega_j^2 \hat{x}_j^2 \right) 
    -\hat{V}\sum_j c_j \hat{x}_j  \nonumber \\
    & = \hat{H}_S + \hat{H}_B + \hat{H}_I.
    \label{eq:H_CL}
\end{align}
Here, we consider heat baths that consist of an infinite number of the harmonic oscillators (bosons), and the momentum, position, mass and frequency of the $j$th oscillator are given by $\hat{p}_j, \hat{x}_j, m_j$ and $\omega_j$, respectively. The system and bath interacts with each other through the system operator $\hat{V}$ and the bath operator $\hat{x}_j$. The quantity $c_j$ is the coupling strength between the system and $j$th bath, and it defines the spectral density $J(\omega)$ as 
\begin{align}
    J(\omega) = \sum_j \frac{c_j^2}{2m_j\omega_j} \delta(\omega-\omega_j).
\end{align}
In the following, we vary the parameters of $J(\omega)$ instead of varying $c_j$'s to change the properties of the heat bath.

\textcolor{black}{The counter term $\hat{H}_c = \hat{V}^2 \sum_j c_j^{2} / (2 m_j \omega_j^2)$ is usually introduced into Eq.~\eqref{eq:H_CL} to compensate for the renormalization of the potential energy~\cite{Weiss2012,Breuer2002,Ingold2002}.
However, we can omit this term in our case because of the following reason.
In this study, we only consider a two-level system for the system $\hat{H}_S$, and the operator $\hat{V}$ is restricted to the Pauli matrices $\hat{\sigma}_\alpha$ ($\alpha \in \{x, y, z\}$).
The equation $\hat{\sigma}_\alpha^2 = \hat{1}$ holds for all the Pauli matrices ($\hat{1}$ is the identity operator of the two-level system), and this indicates that the counter term only shifts the origin of the energy and never affects the dynamics of the system.}

The Liouville--von Neumann equation for the Hamiltonian in Eq.~\eqref{eq:H_CL} in the interaction picture is expressed as
\begin{align}
    \frac{\partial}{\partial t} \tilde{\rho}_\mathrm{tot}(t) = 
    -\frac{i}{\hbar} \tilde{H}_I^\times(t) \tilde{\rho}_\mathrm{tot}(t).
    \label{eq:LvN_I}
\end{align}
The hyper-operator $\hat{O}_1^\times \hat{O}_2 = \hat{O}_1\hat{O}_2-\hat{O}_2\hat{O}_1$ denotes the commutator, and the operators in the interaction picture are given by $\tilde{O}(t) = e^{i(\hat{H}_S+\hat{H}_B)t/\hbar} \hat{O} e^{-i(\hat{H}_S+\hat{H}_B)t/\hbar} = e^{i(\hat{H}_S^\times+\hat{H}_B^\times)t/\hbar}\hat{O}$. Integrating Eq.~\eqref{eq:LvN_I}, we obtain $ \tilde{\rho}_\mathrm{tot}(t) = \tilde{\rho}_\mathrm{tot}(t_0) -i\int_{t_0}^{t} ds \tilde{H}_I^\times(s)\tilde{\rho}_\mathrm{tot}(s) / \hbar$. By substituting this for $\tilde{\rho}_\mathrm{tot}(t)$ on the right-hand side of Eq.~\eqref{eq:LvN_I} and by tracing out the bath degrees of freedom, the equation for the reduced density operator (RDO) of the system is derived as 
\begin{align}
    \frac{\partial}{\partial t} \tilde{\rho}_S(t) = & 
    \frac{\partial}{\partial t} \mathrm{tr}_{B}
    \{\tilde{\rho}_\mathrm{tot}(t)\} \nonumber  \\
    = & -\frac{i}{\hbar} \mathrm{tr}_B\{\tilde{H}_I^\times(t)
    \tilde{\rho}_\mathrm{tot}(t_0)\} \nonumber \\
    & - \frac{1}{\hbar^2}\int_{t_0}^{t} ds
    \mathrm{tr}_B\{
    \tilde{H}_I^\times(t) 
    \tilde{H}_I^\times(s)\tilde{\rho}_\mathrm{tot}(s)\}. \quad
    \label{eq:Master}
\end{align}
Here, $\mathrm{tr}_B$ denotes the partial trace of the heat bath. As the initial states at the time $t = t_0$, we adopt the factorized initial states $\tilde{\rho}_\mathrm{tot}(t_0) = \tilde{\rho}_S(t_0) \otimes \hat{\rho}^{eq}_B$. Here, the equilibrium state of the heat bath is given by $\tilde{\rho}_B(t_0) = \hat{\rho}^{eq}_B = e^{-\beta \hat{H}_B} / \mathrm{tr}\{e^{-\beta \hat{H}_B}\}$, where $\beta = 1 / k_\mathrm{B}T $ is the inverse temperature ($k_\mathrm{B}$ is the Boltzmann constant). Because $\mathrm{tr}_B\{\tilde{x}_j(t)\hat{\rho}^{eq}_B\} = \mathrm{tr}_B\{\hat{\rho}^{eq}_B\tilde{x}_j(t)\} = 0$, the first term in Eq.~\eqref{eq:Master} vanishes.

Equation~\refeq{eq:Master} is analytically exact. In the following, we impose some approximations to obtain equations in the Lindblad form~\cite{Breuer2002,Cohen1992}. First, we consider the \emph{Born approximation}: We assume that the coupling strength between the system and bath is weak. This leads to the approximation in which the total density operator is always factorized, with the density operator of the heat bath time-stationary. Considering the initial states we defined above, the density operator is approximated as $\tilde{\rho}_\mathrm{tot}(s) \simeq \tilde{\rho}_S(s) \otimes \hat{\rho}^{eq}_B$. Due to this approximation, Equation~\refeq{eq:Master} is rewritten as 
\begin{align}
    \frac{\partial}{\partial t} \tilde{\rho}_S(t) = & 
    -\frac{1}{\hbar^2} \int_{t_0}^{t} ds \tilde{V}^\times(t)
    \left\{C(t-s) \tilde{V}(s)\tilde{\rho}_S(s) \right. 
    \nonumber \\
    & \hspace{20ex} \left.
    - C^*(t-s) \tilde{\rho}_S(s)\tilde{V}(s)\right\}, 
    \label{eq:Master+B}
\end{align}
where
\begin{align}
    C(t) & = \hbar \int_{0}^{\infty} d\omega J(\omega)
    \left(\coth \frac{\beta \hbar \omega}{2} \cos \omega t
    - i \sin \omega t\right) \nonumber \\
    & = \int_{-\infty}^{\infty} d\omega S_\beta (\omega)
    e^{-i \omega t}
    \label{eq:CF}
\end{align}
is the two-time correlation function of the heat bath. Here, we have assumed that the spectral density is an odd function, as $J(-\omega) = -J(\omega)$, and defined the spectral noise power as $S_\beta(\omega) = \hbar J(\omega) / (1-e^{-\beta \hbar \omega})$. The function $C(t)$ reaches $0$ as $|t| \to \infty$, and we express the time constant of the decay with $\tau_B$.

To impose another approximation, we introduce a timescale, $\tau_R$, which indicates the timescale of the relaxation dynamics of the system in the interaction picture. We assume that the two-time correlation function $C(t)$ decays fast enough compared with the timescale of the relaxation process of the system, which implies $\tau_B \ll \tau_R$. Due to this approximation, only the integrands at the time $s \simeq t$ contribute to the integration of Eq.~\eqref{eq:Master+B}, and furthermore, we can assume $\tilde{\rho}_S(s) \simeq \tilde{\rho}_S(t)$ at the time $s \simeq t$. With this approximation, which is referred to as the \emph{Markov approximation}, we can replace $\tilde{\rho}_S(s)$ of the integrand in Eq.~\eqref{eq:Master+B} by $\tilde{\rho}_S(t)$: The equation reads
\begin{align}
    \frac{\partial}{\partial t} \tilde{\rho}_S(t) = & 
    -\frac{1}{\hbar^2} \int_{-\infty}^{t} ds \tilde{V}^\times(t)
    \left\{C(t-s) \tilde{V}(s)\tilde{\rho}_S(t) \right. 
    \nonumber \\
    & \hspace{20ex} \left.
    - C^*(t-s) \tilde{\rho}_S(t)\tilde{V}(s)\right\}.
    \label{eq:Master+BM}
\end{align}
Here, we take the limit $t_0 \to -\infty$ to remove the dependence on the initial time. This approximation is also based on the fact that $C(t)$ decays sufficiently fast. We note, however, that this timescale separation does no longer exist at very low temperatures when reservoir correlations decay algebraically rather than exponentially in time. For example, for Ohmic reservoirs $J(\omega)\propto \omega$, one has at $T=0$ that $C(t)\propto 1/t^2$.

Equation~\refeq{eq:Master+BM} only depends on the density operator at the time $t$, and in this sense, it is a time-local equation. However, this is not in the Lindblad form; in the following subsections, we impose further approximation and demonstrate how this equation is transformed into Lindblad equations.

\subsection{Lindblad equation: rotating wave approximation}
In this subsection, we follow the standard method~\cite{Breuer2002} to obtain the LE from Eq.~\eqref{eq:Master+BM}. We decompose the operator $\hat{V}$ as
\begin{align}
    \hat{V} & = \sum_{\omega} 
    \sum_{\varepsilon'-\varepsilon=\hbar\omega} 
    \ketbra{\varepsilon}{\varepsilon}\hat{V}
    \ketbra{\varepsilon'}{\varepsilon'} \nonumber \\
    & = \sum_{\omega} \hat{V}(\omega),
    \label{eq:V_decomp}
\end{align}
where $\ket{\varepsilon}$ is the eigenvector of the system Hamiltonian $\hat{H}_S\ket{\varepsilon} = \varepsilon\ket{\varepsilon}$. The interaction picture of $\hat{V}(\omega)$ is given by $e^{i\hat{H}_S^\times t/\hbar} \hat{V}(\omega) = e^{-i\omega t} \hat{V}(\omega)$. By using this, we obtain Eq.~\eqref{eq:Master+BM} in the frequency-dependent form as 
\begin{align}
    \frac{\partial}{\partial t} \tilde{\rho}_S(t) = &
    \frac{1}{\hbar^2} \sum_{\omega, \omega'}
    [e^{-i(\omega'-\omega)t} \Gamma(\omega)
    \begin{aligned}[t]
        \Bigl(&\hat{V}(\omega) \tilde{\rho}_S(t)
        \hat{V}^{\dagger}(\omega') \\
        & - \hat{V}^{\dagger}(\omega')\hat{V}(\omega)
        \tilde{\rho}_S(t)\Bigr)
    \end{aligned} \nonumber \\
    & \hspace{8ex} + \mathrm{H. c.}] \label{eq:Master_freq},
\end{align}
where H.c. denotes the Hermitian conjugates and $\Gamma(\omega)$ is defined as follows:
\begin{align}
    \Gamma(\omega) & = \int_{0}^{\infty} ds C(s) e^{i \omega s}
    \nonumber \\
    & = \int_{0}^{\infty}ds \int_{-\infty}^{\infty} d\omega'
    S_\beta(\omega') e^{i (\omega - \omega') s} \nonumber \\
    & = \pi S_\beta(\omega)
    + i\mathcal{P}\int_{-\infty}^{\infty} d\omega' 
    \frac{S_\beta(\omega')}{\omega-\omega'} \nonumber \\
    & = \frac{1}{2}\gamma(\omega) + i \Lambda(\omega). 
\end{align}
Here, the notation $\mathcal{P}$ is the Cauchy principal value.

To obtain an equation in the Lindblad form, we impose the \emph{rotating wave approximation} (RWA): Because we consider the slow dynamics of $\tilde{\rho}_S(t)$, the contributions of the fast oscillating terms $e^{\pm i(\omega'-\omega)t}$ with $|\omega'-\omega| \gg 1/\tau_R$ are negligible. Therefore we only consider the term with $\omega = \omega'$ in Eq.~\eqref{eq:Master_freq}, and obtain
\begin{align}
    \frac{\partial}{\partial t} \hat{\rho}_S(t) = & 
    -\frac{i}{\hbar}\left(\hat{H}_S^\times
    +\hat{\mathcal{H}}_\mathrm{LS}^\times\right)
    \hat{\rho}_S(t) \nonumber \\
    & + \frac{1}{\hbar^2} \sum_{\omega} 
    \gamma(\omega)\left[\hat{V}(\omega) \hat{\rho}_S(t)
    \hat{V}^\dagger(\omega) \right. \nonumber \\
    & \hspace{16ex}\left.
    - \frac{1}{2}\left(\hat{V}^\dagger(\omega)\hat{V}(\omega)\right)^\circ
    \hat{\rho}_S(t)\right], \quad
    \label{eq:LE}
\end{align}
which is in the Lindblad form. Here, we return to the Schr\"odinger picture, $\hat{\rho}_S(t) = e^{-i\hat{H}_S^\times t /\hbar}\tilde{\rho}_S(t)$, and the hyper-operator $\hat{O}_1^\circ \hat{O}_2 = \hat{O}_1\hat{O}_2 + \hat{O}_2\hat{O}_1$ is the anticommutator. We introduce the Lamb-shift Hamiltonian as $\hat{\mathcal{H}}_\mathrm{LS} = \sum_{\omega} \Lambda(\omega) \hat{V}^\dagger(\omega) \hat{V}(\omega) / \hbar$.

\subsection{Universal Lindblad equation: decomposition of the spectral density}
In the preceding subsection, we consider the RWA, which depends on the properties of the system, to obtain the LE. A recent study~\cite{NathanPRB2020} demonstrated that we can derive an equation in the Lindblad form irrespective of the properties of the system, which is referred to as the universal Lindblad equation (ULE). Here, we briefly review the derivation of the ULE; for the details, see Ref.~\onlinecite{NathanPRB2020}.

We consider the square root of the spectral noise power, which is defined as
\begin{align}
     g(\omega) = \sqrt{\frac{S_\beta(\omega)}{2\pi}},
     \label{eq:JumpCorr}
\end{align}
and its Fourier transform $g(t) = \int_{-\infty}^{\infty} g(\omega) e^{-i\omega t}$, which is referred to as the ``jump correlator'' in the original paper~\cite{NathanPRB2020}. To obtain the ULE, we exploit the properties of $g(t)$ instead of $C(t)$.
With the relation Eq.~\eqref{eq:JumpCorr}, the two-time correlation function is rewritten as $C(t-t') = \int_{-\infty}^{\infty} ds g(t-s)g(s-t')$. Substituting this in Eq.~\eqref{eq:Master+BM}, we obtain
\begin{align}
    \frac{\partial}{\partial t} \tilde{\rho}_S(t) = & 
    \int_{-\infty}^{\infty} dt' \int_{-\infty}^{\infty} ds
    \mathcal{F}(t, s, t')[\tilde{\rho}_S(t)].
    \label{eq:ULE_I0}
\end{align}
Here, we have introduced $\mathcal{F}(t, s, t')[\tilde{\rho}_S(t)]$ as 
\begin{align}
    & \mathcal{F}(t, s, t')[\tilde{\rho}_S(t)] \nonumber \\
    = & -\frac{1}{\hbar^2} \theta(t-t') \Bigl(g(t-s)g(s-t') 
    \tilde{V}^\times(t)\tilde{V}(t')\tilde{\rho}_S(t)
    \nonumber \\
    & \hspace{15ex}
    - g^*(t-s)g^*(s-t')\tilde{V}^\times(t)\tilde{\rho}_S(t)
    \tilde{V}(t') \Bigr),
    \label{eq:K_ULE}
\end{align}
where $\theta(t)$ is the Heaviside step function.

Similar to $C(t)$, the jump correlator $g(t)$ is the function that decays fast to $0$ as $|t|$ grows when the Born--Markov approximation is imposed.
\textcolor{black}{This indicates that only the integrand around the region $t \simeq t' \simeq s$ contributes to the evaluation of the integral in Eq.~\eqref{eq:ULE_I0}.
By utilizing this, we can change $t$ with $s$ in Eq.~\eqref{eq:ULE_I0}, and it is rewritten as}
\begin{align}
    \frac{\partial}{\partial t} \tilde{\rho}_S(t) = & 
    \int_{-\infty}^{\infty} ds \int_{-\infty}^{\infty} ds'
    \mathcal{F}(s, t, s')[\tilde{\rho}_S(t)].
    \label{eq:ULE_I}
\end{align}
\textcolor{black}{Note that in terms of the accuracy of the approximation, the replacement of the density operator $\tilde{\rho}_S(t)$ with $\tilde{\rho}_S(s)$ in Eq.~\eqref{eq:ULE_I0} is equivalent to the conventional Markov approximation~\cite{NathanPRB2020}, in which we replace $\tilde{\rho}_S(s)$ to $\tilde{\rho}_S(t)$ in Eq.~\eqref{eq:Master+B} to obtain Eq.~\eqref{eq:Master+BM}.}

By substituting Eq.~\eqref{eq:K_ULE} into Eq.~\eqref{eq:ULE_I} and return to the Schr\"odinger picture, we obtain the ULE as
\begin{align}
    \frac{\partial}{\partial t} \hat{\rho}_S(t) = & 
    -\frac{i}{\hbar}\left(\hat{H}_S^\times
    +\hat{\mathscr{H}}_\mathrm{LS}^\times\right)
    \hat{\rho}_S(t) \nonumber \\
    & + \frac{1}{\hbar^2} 
    \left[\hat{L} \hat{\rho}_S(t)
    \hat{L}^\dagger 
    - \frac{1}{2}\left(\hat{L}^\dagger\hat{L}\right)^\circ
    \hat{\rho}_S(t)\right].
    \label{eq:ULE}
\end{align}
The Lindblad operator $\hat{L}$ is given by
\begin{align}
    \hat{L} & = e^{-i\hat{H}_S^\times t/\hbar}
    \int_{-\infty}^{\infty} ds g(t-s)
    \tilde{V}(s) \nonumber \\
    & = \int_{-\infty}^{\infty} ds g(s) 
    \tilde{V}(-s),
    \label{eq:L_ULE}
\end{align}
and the Lamb-shift term is expressed as
\begin{align}
    \hat{\mathscr{H}}_\mathrm{LS} & = 
    \begin{aligned}[t]
        & \frac{1}{2i\hbar}
        e^{-i\hat{H}_S^\times t/\hbar} 
        \int_{-\infty}^{\infty}\hspace{-1.5ex}ds
        \int_{-\infty}^{\infty}\hspace{-1.5ex}ds'
        \\
        & \times
        \mathrm{sgn}(s-s')g(s-t)g(t-s')\tilde{V}(s)\tilde{V}(s')
    \end{aligned}
    \nonumber \\
    & = \frac{1}{2i\hbar}
    \int_{-\infty}^{\infty}\hspace{-1.5ex}ds
    \int_{-\infty}^{\infty}\hspace{-1.5ex}ds'
    \mathrm{sgn}(s-s')g(s)g(-s')\tilde{V}(s)\tilde{V}(s'),
    \label{eq:LS_ULE}
\end{align}
where $\mathrm{sgn}(s)$ is the sign function.

When we expand $\hat{L}$ and $\hat{\mathscr{H}}_\mathrm{LS}$ on the basis of the eigenvectors of the system Hamiltonian $\ket{n}$, where $\hat{H}_S\ket{n} = \hbar \omega_n \ket{n}$, Equations~\refeq{eq:L_ULE} and \refeq{eq:LS_ULE} are given by
\begin{gather}
    \hat{L} = \sum_{m, n} \sqrt{2\pi S_\beta(\omega_{nm})}
    V_{mn} \ketbra{m}{n}, \label{eq:L_ULE1}\\
    \hat{\mathscr{H}_\mathrm{LS}} = \frac{1}{\hbar}
    \sum_{m, n, l} V_{ml}V_{ln} f(\omega_{lm}, \omega_{nl})
    \ketbra{m}{n},
\end{gather}
where $\omega_{nm} = \omega_n - \omega_m$, $V_{mn} = \mel{m}{\hat{V}}{n}$ and 
\begin{align}
    f(\omega_1, \omega_2) = - 2\pi \mathcal{P}
    \int_{-\infty}^{\infty} d\omega
    \frac{g(\omega-\omega_1)g(\omega+\omega_2)}{\omega}.
\end{align}

The ULE is derived on the basis of the Eq.~\eqref{eq:Master+BM}: The starting points to obtain both LE and ULE are same. The main difference between the LE and ULE is that in the LE, we impose the RWA, while we do not in the ULE. We can derive an equation in the Lindblad form by only utilizing the properties of the heat bath.

In Sec.~\ref{sec:results}, we see the differences between the standard LE and ULE through examples of a two-level system. 

\section{Derivation of the Hierarchical Equations of Motion} \label{sec:deriveHEOM}
In this section, we derive the hierarchical equations of motion (HEOM)~\cite{heomHigh}, particularly its recent extension to the FP-HEOM~\cite{XuPRL2022}, which describes the open quantum systems in a numerically rigorous manner.

In the same way as the derivation of the LEs, we consider the Caldeira--Leggett Hamiltonian as the model. The density operator at the time $t$ is described in the path-integral form, and by tracing out the bath degrees of freedom, the RDO of the systems is expressed as
\begin{align}
    & \mel{\alpha} {\hat{\rho}_S(t)}{\alpha'} \nonumber \\
    = & \int\frac{d\alpha_i d\alpha'_i}{\mathscr{N}^2}
    \int_{\alpha(t_0)=\alpha_i}^{\alpha(t)=\alpha}
    \hspace{-5ex}\mathcal{D}[\alpha(\cdot)]
    \int_{\alpha'(t_0)=\alpha'_i}^{\alpha'(t)=\alpha'}
    \hspace{-5ex}\mathcal{D}[\alpha'(\cdot)] \nonumber \\
    & \times 
    e^{iS_S[\dot{\alpha}, \alpha; t]} 
    \mel{\alpha_i}{\hat{\rho}_S(t_0)}
    {\alpha'_i} e^{-iS_S[\dot{\alpha}', \alpha';t]}
    \mathscr{F}[\alpha, \alpha';t]. \qquad
    \label{eq:RDO_HEOM}
\end{align}
Here, $\ket{\alpha}$ is the ket vector with the boson-coherent, fermion-coherent, spin-coherent and displacement representation. The normalization factor $\mathscr{N}$ depends on the representation of $\alpha$. We consider the factorized initial states again. The quantity $S_S[\dot{\alpha}, \alpha; t]$ is the action of the system.

The functional $\mathscr{F}[\alpha, \alpha'; t]$ is referred to as the influence functional and is given by
\begin{align}
    & \mathscr{F}[\alpha, \alpha'; t] \nonumber \\
    = & \exp\Biggl[-\frac{1}{\hbar^2} \int_{t_0}^{t} dt'
    \int_{t_0}^{t'} dt''V^\times(\alpha, \alpha'; t')
    \nonumber \\
    & \times \left\{C(t'-t'')V(\alpha; t'')
    - C^{*}(t'-t'')V(\alpha'; t'')
    \right\}\Biggr]. \qquad
    \label{eq:IF}
\end{align}
Here we have introduced the path-integral representations of the operators and hyperoperators as $V(\alpha; t)$ and $V^\times(\alpha, \alpha'; t) = V(\alpha; t) - V(\alpha'; t)$, respectively. Note that the time derivative of Eq.~\eqref{eq:IF} has a close relation to the right-hand side of Eq.~\eqref{eq:Master+B}. 

The FP-HEOM now uses an  expansion of the two-time correlation function as $C(t) = \sum_{k=1}^{K} d_k e^{-z_k t}$ for the time $t > 0$. The coefficients $d_k$ and $z_k$ are complex numbers, and $\mathrm{Re}\{z_k\} > 0$. By expressing $S_\beta(\omega)$ in the barycentric representation~\cite{XuPRL2022}, we can obtain the set of $\{d_k\}$ and $\{z_k\}$ with high accuracy and such that the number of coefficients $K$ remains small to moderate for almost all spectral bath densities, over the whole temperature range down to $T=0$, and arbitrary coupling strengths.

To obtain equations of motion for the RDO in Eq.~\eqref{eq:RDO_HEOM} without any additional approximations, we introduce a new quantity, ``auxiliary density operator'' (ADO), which is in the form of
\begin{align}
    & \mel{\alpha}{\hat{\rho}_{\vec{m}, \vec{n}}(t)}
    {\alpha'}\nonumber \\
    = & \int\frac{d\alpha_i d\alpha'_i}{\mathscr{N}^2}
    \int_{\alpha(t_0)=\alpha_i}^{\alpha(t)=\alpha}
    \hspace{-5ex}\mathcal{D}[\alpha(\cdot)]
    \int_{\alpha'(t_0)=\alpha'_i}^{\alpha'(t)=\alpha'}
    \hspace{-5ex}\mathcal{D}[\alpha'(\cdot)] \nonumber \\
    & \times 
    \prod_{k=1}^{K}
    \begin{aligned}[t]
        & \left(\int_{t_0}^{t}dt'' d_k e^{-z_k(t-t'')}
        \frac{-i}{\hbar}V(\alpha; t'')\right)^{m_k} \\
        & \times \left(\int_{t_0}^{t}dt'' d^*_k e^{-z^*_k(t-t'')}
        \frac{i}{\hbar} V(\alpha'; t'')\right)^{n_k}
    \end{aligned} \nonumber \\
    & \times 
    e^{iS_S[\dot{\alpha}, \alpha; t]} 
    \mel{\alpha_i}{\hat{\rho}_S(t_0)}
    {\alpha'_i} e^{-iS_S[\dot{\alpha}', \alpha';t]}
    \mathscr{F}[\alpha, \alpha';t].
\end{align}
Here, we have introduced the vectors $\vec{m} = [m_1, \ldots, m_K]$ and $\vec{n} = [n_1, \ldots, n_K]$, whose elements are non-negative integers, to distinguish ADOs. The ADO $\hat{\rho}_{\vec{0}, \vec{0}}(t)$ corresponds to the RDO.

Considering the time derivative,
\begin{widetext}
    \begin{align}
        & \frac{\partial \hat{\rho}_{\vec{m}, \vec{n}}(t)}
        {\partial t}
        = \int 
        \frac{d\alpha d\alpha'}{\mathscr{N}^2} \ket{\alpha}
        \lim_{\Delta t \to 0} 
        \frac{\mel{\alpha}
        {\hat{\rho}_{\vec{m}, \vec{n}}(t+\Delta t)}{\alpha'}
        - \mel{\alpha}{\hat{\rho}_{\vec{m}, \vec{n}}(t)}
        {\alpha'}}
        {\Delta t}
        \bra{\alpha'},
    \end{align}
\end{widetext}
we obtain the following equations of motion:
\begin{align}
    \frac{\partial}{\partial t}
    \hat{\rho}_{\vec{m}, \vec{n}}(t) = & 
    -\frac{i}{\hbar} \hat{H}^\times_S
    \hat{\rho}_{\vec{m}, \vec{n}}(t)
    - \sum_{k=1}^{K} (m_k z_k+n_k z^*_k)
    \hat{\rho}_{\vec{m}, \vec{n}}(t) \nonumber \\
    & -\frac{i}{\hbar} \hat{V}^\times
    \sum_{k=1}^{K}
    \left(\hat{\rho}_{\vec{m}+\vec{e}_k, \vec{n}}(t) 
    + \hat{\rho}_{\vec{m}, \vec{n}+\vec{e}_k}(t)\right)
    \nonumber \\
    & + \sum_{k=1}^{K}\biggl(-m_k d_k \frac{i}{\hbar} \hat{V}
    \hat{\rho}_{\vec{m}-\vec{e}_k, \vec{n}}(t) \nonumber \\
    & \hspace{8ex} + n_k d^*_k \frac{i}{\hbar}
    \hat{\rho}_{\vec{m}, \vec{n}-\vec{e}_k}(t)\hat{V}\biggr),
    \label{eq:HEOM}
\end{align}
which we refer to as the HEOM.
Here $\vec{e}_k$ is the unit vector of the $k$th element. In the following, we choose the initial states of ADOs as $\hat{\rho}_{\vec{0}, \vec{0}}(t_0) = \hat{\rho}_S(t_0)$ and $\hat{\rho}_{\vec{m}\neq\vec{0}, \vec{n}\neq\vec{0}}(t_0) = 0$.

Here, we remark on the difference between the Lindblad equations and HEOM. Considering the case of $\vec{m}=\vec{n}=\vec{0}$ in Eq.~\eqref{eq:HEOM}, the time differential equation for the RDO is written as
\begin{align}
    \frac{\partial}{\partial t}
    \hat{\rho}_{\vec{0}, \vec{0}}(t) = & 
    -\frac{i}{\hbar} \hat{H}^\times_S
    \hat{\rho}_{\vec{0}, \vec{0}}(t) \nonumber \\
    & -\frac{i}{\hbar} \hat{V}^\times
    \sum_{k=1}^{K} \left(\hat{\rho}_{\vec{e}_k, \vec{0}}(t) 
    + \hat{\rho}_{\vec{0}, \vec{e}_k}(t)\right).
\end{align}
The Born--Markov approximation corresponds to the approximation of the second term $\hat{\rho}_{\vec{e}_k, \vec{0}}(t)$ and $\hat{\rho}_{\vec{0}, \vec{e}_k}(t)$ by means of  $\hat{\rho}_{\vec{0}, \vec{0}}(t)$. Due to the introduction of the ADOs, we can express the equations of motion without the approximations. On the other hand, the number of equations increases, and more computational resources are needed.

To obtain the closed set of the simultaneous differential equations, we need to truncate Eq.~\eqref{eq:HEOM}: we define the depth of the hierarchy as $\mathcal{N} = \sum_{k=1}^{K} (m_k+n_k)$, and always set $\hat{\rho}_{\vec{m}, \vec{n}}(t) = 0$ for the ADOs with $\mathcal{N} > \mathcal{N}_{\max}$. In the following calculations, we vary the value $\mathcal{N}_{\max}$ and confirm that the dynamics of RDO converges.

To obtain HEOM in Eq.~\eqref{eq:HEOM}, we do not consider any approximations, except for the form of the Hamiltonian, in which the heat bath is represented by the harmonic oscillators. In the following section, we use results obtained with HEOM as references for those obtained with the LE and ULE.

\subsection*{Remarks on the influence functional and Markovianity}
If we assume that the two-time correlation function of the heat bath is given by Dirac delta function $C(t) \simeq \gamma \delta(t)$ ($\gamma > 0$), Eq.~\refeq{eq:IF} is rewritten as 
\begin{align}
    & \mathscr{F}[\alpha, \alpha; t] \nonumber \\
    \simeq & \exp\left[
    -\frac{\gamma}{2\hbar^2}\int_{t_0}^{t} \!\!\!dt'
    V^\times(\alpha, \alpha'; t')\left\{V(\alpha; t')
    -V(\alpha'; t')\right\}\right] \nonumber \\
    = & \exp\left[
    \frac{\gamma}{\hbar^2}\int_{t_0}^{t} \!\!\!dt'
    \left\{V(\alpha;t') V(\alpha';t')
    - \frac{1}{2} (V^2)^\circ(\alpha, \alpha';t')\right\}
    \right],
    \label{eq:IF_delta}
\end{align}
where $(V^2)^\circ(\alpha, \alpha';t) = V(\alpha;t)V(\alpha;t) + V(\alpha';t)V(\alpha';t)$.
\textcolor{black}{The time derivative of Eq.~\eqref{eq:RDO_HEOM} with this influence functional is expressed as the following equation:
\begin{align}
    \frac{\partial}{\partial t} \hat{\rho}_S(t)
    \!= & -\!\frac{i}{\hbar} \hat{H}_S^\times \hat{\rho}_S(t)
    + \frac{\gamma}{\hbar^2}\left[\hat{V}\hat{\rho}_S(t)\hat{V}
    - \frac{1}{2} (\hat{V}^{2})^\circ \hat{\rho}_S(t)\right],
    \label{eq:LE_Delta}
\end{align}
which is in the Lindblad form.
It is worth noting that the LE [Eq.~\eqref{eq:LE}], ULE [Eq.~\eqref{eq:ULE}] and Eq.~\eqref{eq:LE_Delta} do not coincide generally, considering that the coefficients and operators in those equations are not same.}
Equation~\refeq{eq:IF_delta} holds for any values of $\gamma$~\cite{Breuer2002}. This contrasts with the LE and ULE, because the Born--Markov approximation is based on the assumption that the coupling strength between the system and bath is weak~\cite{Breuer2002, Cohen1992}. Following this line of argument, the \emph{singular coupling limit} has been introduced to recover the Markovianity in previous studies~\cite{Breuer2002,Rivas2012,Alicki1987}.
\textcolor{black}{The Caldeira--Leggett master equation~\cite{CaldeiraPA1983,DiosiEPL1993} is also based on the similar argument to the above, but the form of this equation is not same as Eq.~\eqref{eq:IF_delta} and is not the Lindblad form:} With the higher-order correction with respect to the inverse temperature, we obtain equations in the Lindblad form~\cite{Breuer2002, DiosiPA1993}.

\textcolor{black}{When we consider the situation in which the system Hamiltonian and the system part of the system--bath interaction commute, $\hat{H}_S^\times \hat{V} = 0$, we can rigorously express the RDO without ADOs. The time derivative of the RDO is similar to Eq.~\eqref{eq:LE_Delta}, but the coefficient $\gamma$ is time-dependent in this case (also, a time-dependent Lamb-shift term may be included)~\cite{DollCP2008}.
Similar to the time-convolutionless (TCL) master equation, non-Markovian effects can be expressed in this case due to the time-dependent coefficients.}

\section{Numerical Results} \label{sec:results}
In this section, we illustrate how the approximations introduced above cause errors in the numerical simulations by comparing the results obtained by the LE, ULE, and FP-HEOM.

\subsection{Model: qubit dynamics}
As a test case, we adopt a two-level system (TLS) for the system. The operators $\hat{H}_S$ and $\hat{V}$ are respectively given by
\begin{align}
    \hat{H}_S & = \frac{\hbar\omega_q}{2} \hat{\sigma}_z, 
    & \hat{V} & = \hbar \hat{\sigma}_x, \label{eq:TLS}
\end{align}
where $\hat{\sigma}_\alpha$ ($\alpha \in \{x, y, z\}$) is the Pauli matrix. We express the ground state and excited state as $\ket{0}$ and $\ket{1}$. Using the relations $\hat{H}_S\ket{0} = -\hbar \omega_q \ket{0} / 2$ and $\hat{H}_S \ket{1} = \hbar \omega_q \ket{1} / 2$, the decomposition of $\hat{V}$ in Eq.~\eqref{eq:V_decomp} is expressed as 
\begin{align}
    \hat{V} & = \hat{V}(\omega_q) + \hat{V}(-\omega_q)
    \nonumber \\
    & = \hat{V}^\dagger(-\omega_q) + \hat{V}^\dagger(\omega_q)
    \nonumber \\
    & = \hbar \hat{\sigma}_{-} + \hbar \hat{\sigma}_{+}.
\end{align}
Here, $\hat{\sigma}_{\pm} = (\hat{\sigma}_x \pm i \hat{\sigma}_y)/2$ is the raising and lowering operator of the $1/2$-spin.

The Lindblad operator for the ULE in Eq.~\eqref{eq:L_ULE1} is given by
\begin{align}
    \hat{L} = \hbar \sqrt{2\pi S_\beta(\omega_q)} 
    \hat{\sigma}_{-} 
    + \hbar \sqrt{2\pi S_\beta(-\omega_q)} \hat{\sigma}_{+},
\end{align}
and the Lamb-shift Hamiltonians for the LE and ULE are as follows:
\begin{align}
    \hat{\mathcal{H}}_\mathrm{LS} & = 
    \hbar \Lambda(\omega_q) \ketbra{1}{1}
    + \hbar \Lambda(-\omega_q) \ketbra{0}{0}, \\
    \hat{\mathscr{H}}_\mathrm{LS} & = 
    \hbar f(-\omega_q, \omega_q) \ketbra{1}{1} 
    + \hbar f(\omega_q, -\omega_q) \ketbra{0}{0}.
\end{align}
Note that in our case, both Lamb shifts of the LE and ULE coincide, which is indicated as 
\begin{align}
    f(\mp\omega_q, \pm\omega_q) = & -2\pi \mathcal{P}
    \int_{-\infty}^{\infty} d\omega
    \frac{g^2(\omega \pm \omega_q)}{\omega} \nonumber \\
    = & - \mathcal{P}\int_{-\infty}^{\infty} d\omega
    \frac{S_\beta(\omega \pm \omega_q)}{\omega} \nonumber \\
    = & -\mathcal{P}\int_{-\infty}^{\infty} d\omega
    \frac{S_\beta(\omega)}{\omega \mp \omega_q} \nonumber \\
    = & \Lambda(\pm \omega_q).
    \label{eq:LS_TLS}
\end{align}
Introducing the matrix elements of the RDO defined as $\mel{i}{\hat{\rho}_S(t)}{j} = \rho_{ij}(t)$ $(i, j = 0, 1)$, we obtain the differential equations for each element. The LE and ULE for the diagonal elements coincide, which is in the form of 
\begin{align}
    \begin{bmatrix}
        \dot{\rho}_{00}(t) \\
        \dot{\rho}_{11}(t)
    \end{bmatrix}
    = 2\pi
    \begin{bmatrix}
        -S_\beta(-\omega_q) & +S_\beta(\omega_q) \\
        +S_\beta(-\omega_q) & -S_\beta(\omega_q)
    \end{bmatrix}
    \begin{bmatrix}
        \rho_{00}(t) \\
        \rho_{11}(t)
    \end{bmatrix}, \label{eq:TLS_diagonal}
\end{align}
while the equations for the off-diagonal elements are different as follows. For the LE in Eq.~\eqref{eq:LE}, we obtain
\begin{align}
    \left\{
    \begin{aligned}
        \dot{\rho}_{01}(t) = &\left(+i \tilde{\omega} 
        - \frac{\gamma_\mathrm{r}}{2}\right)
        \rho_{01}(t) \\
        \dot{\rho}_{10}(t) = &\left(-i \tilde{\omega}
        - \frac{\gamma_\mathrm{r}}{2}\right)
        \rho_{10}(t)
    \end{aligned}
    \right. ,\label{eq:LE_TLS}
\end{align}
and for the ULE in Eq.~\eqref{eq:ULE}, we obtain
\begin{align}
    \left\{
    \begin{aligned}
        \dot{\rho}_{01}(t) \!=\! &\left(+i \tilde{\omega} 
        - \frac{\gamma_\mathrm{r}}{2} \right)
        \rho_{01}(t)
        + \Delta \rho_{10}(t) \\
        \dot{\rho}_{10}(t) \!=\! &\left(-i \tilde{\omega}
        -\frac{\gamma_\mathrm{r}}{2} \right)
        \rho_{10}(t)
        + \Delta \rho_{01}(t)
    \end{aligned}
    \right. .\label{eq:ULE_TLS}
\end{align}
Here, the frequency $\tilde{\omega}$ is defined as $\tilde{\omega} = \omega_q + \Lambda(\omega_q) - \Lambda(-\omega_q)$, and the quantity
\begin{align}
    \gamma_\mathrm{r} = 2\pi (S_\beta(\omega_q) + S_\beta(-\omega_q))
\end{align}
is introduced. The term $\Delta = 2\pi \sqrt{S_\beta(\omega_q)S_\beta(-\omega_q)}$ only appears in the ULE.
\textcolor{black}{Due to this term, the effective frequency of the Larmor precession is changed from $\tilde{\omega}$ to $\sqrt{\tilde{\omega}^2-\Delta^2}$ in the ULE case.}

Diagonalizing the matrix in Eq.~\eqref{eq:TLS_diagonal}, one obtains two eigenvalues, namely,  the eigenvalue $0$, and $\gamma_\mathrm{r}$. The vanishing eigenvalue corresponds to the equilibrium state while the rate $\gamma_\mathrm{r}$ describes the monoexponential relaxation process. Note that in the LE, the decay rate of the decoherence (dephasing) is a half of that of the population relaxation when the system part of the system--bath coupling is given by $\hat{V} = \hbar \hat{\sigma}_x$.  From the definition of the spectral noise power, the fluctuation-dissipation relation, $S_\beta(-\omega_q) = e^{-\beta\hbar\omega_q} S_\beta(\omega_q)$, is derived, here also known as detailed balance. Applying this relation to the time-independent eigenvector, we obtain the equilibrium distribution of the states $\ket{0}$ and $\ket{1}$ which in the LE is given by the Boltzmann distribution with respect to the bare system Hamiltonian $\hat{H}_S$, as $\rho_{jj}^{eq, \mathrm{LE}} = e^{(-1)^j\beta\hbar\omega_q/2}/[2\cosh(\beta\hbar\omega_q/2)]$ $(j = 0, 1)$.

\textcolor{black}{The difference of the off-diagonal elements between the LE and ULE is caused by the finite $\Delta$ in Eq.~\eqref{eq:ULE_TLS}, and it vanishes when we consider the zero-temperature limit.}
This is because we can evaluate the value $S_\beta(-\omega_q)$ as $\lim_{\beta \to \infty} S_{\beta}(-\omega_q) = 0$.

In the high-temperature limit $\beta \to 0$, the equation $S_\beta(\omega) = S_\beta(-\omega)$ holds, and therefore the relation $\gamma_\mathrm{r} / 2 = \Delta$ is derived. With this relation, it is demonstrated that Eq.~\refeq{eq:ULE_TLS} is same as Eq.~\eqref{eq:LE_Delta} in this limit, except for the difference of the coefficients [$\omega_q$ and $\gamma$ in Eq.~\eqref{eq:LE_Delta} and $\tilde{\omega}$ and $\gamma_r / 2 = \Delta$ in Eq.~\eqref{eq:ULE_TLS}].
By contrast, LE generally does not correspond to Eq.~\eqref{eq:LE_Delta} due to the lack of the $\Delta$-term in Eq.~\eqref{eq:LE_TLS}.
\textcolor{black}{Note that although the Caldeira--Leggett master equation (CLM) is derived in the high-temperature limit, it does neither coincide with the ULE nor the LE in this limit, as Eq.~\eqref{eq:LE_Delta} has a different structure as the CLM. The CLM is derived by additionally assuming a qubit--reservoir coupling larger or on the order of typical system frequencies (Brownian motion limit), while the LE and ULE require the opposite (quantum optical limit).}
If we neglect the term originated from the imaginary part of $C(t)$ in the Caldeira--Leggett master equation, which appears to be small in the high-temperature limit, the ULE, Eq.~\eqref{eq:LE_Delta} and Caldeira--Leggett master equation coincide.

To conduct numerical calculations, particularly within the exact HEOM approach, we chose the following spectral density:
\begin{align}
    J(\omega) = \mathrm{sgn}(\omega)
    \frac{\kappa \omega_\mathrm{ph}^{1-s}
    |\omega|^{s}}{(1+(\omega/\omega_c)^2)^2} .
    \label{eq:SD}
\end{align}
Here, the quantities $\kappa$ and $\omega_c$ are the coupling strength between the system and bath and cutoff frequency, respectively. The quantity $\omega_\mathrm{ph}$ has been introduced to fix the unit of $\kappa$ irrespective of the exponent $s$. Note that the ratio of the decay rate to the system frequency, $\gamma_\mathrm{r}/\omega_q$, is proportional to $2\pi\hbar\kappa$, which is a dimensionless quantity. We set $\omega_\mathrm{ph}$ to $\omega_q$: With this parameter value, we obtain the value $\gamma_\mathrm{r} / \omega_q = 2 \pi \hbar \kappa \coth(\beta \hbar \omega_q /2) / (1+ (\omega_q/\omega_c)^2)^2$, which is independent of the exponent $s$. We set $\omega_q$ as the unit of the frequency, and chose parameter values as $\beta\hbar\omega_q = 5$, $\omega_c/\omega_q = 50$, and $2\pi\hbar\kappa = 10^{-3}$, which seems to be in the region where the LE and ULE can be applied.
\textcolor{black}{In addition, we consider the parameter value $2\pi\hbar\kappa = 10^{-2}$ to study the violation of the weak-coupling approximation.}

For the spectral density, we consider two values of $s$. In one case, we chose $s = 1$, which corresponds to the Ohmic spectral density. The Ohmic spectral density has been widely adopted for the studies of open quantum systems, because the classical limit of the Ohmic spectral density with $\omega_c \to \infty$ leads to the Langevin equation. In the other case, we chose $s=1/4$: The spectral density with the condition $s < 1$ is referred to as the sub-Ohmic spectral density. It is suggested that the transmon qubit is subject to $1/f^\varepsilon$ $(\varepsilon > 0)$ noise~\cite{IthierPRB2005,BylanderNP2011}, and the sub-Ohmic spectral density exhibits $1/f^\varepsilon$ behavior in the low-frequency region, as $S_\beta(\omega) \propto \kappa k_\mathrm{B}T /\omega^{1-s}$.
For the calculations of HEOM with the coupling strength $2\pi\hbar\kappa = 10^{-3}$, we computed the set $\{d_k\}$ and $\{z_k\}$ with $K=15$ for the Ohmic spectral density, and $K=30$ for the sub-Ohmic spectral density. We chose $\mathcal{N}_{\max} = 2$ and $3$ for the maximum depth of the hierarchy with the Ohmic and sub-Ohmic spectral density, respectively.
\textcolor{black}{For the stronger coupling case, $2\pi\hbar\kappa=10^{-2}$, the parameter values for the FP-HEOM calculations are as follows: $K = 12$ and $\mathcal{N}_{\max} = 2$ for the Ohmic case, and $K = 25$ and $\mathcal{N}_{\max} = 4$ for the sub-Ohmic case.}

For the frequencies of the Lamb shift in Eq.~\eqref{eq:LS_TLS}, we consider the Ohmic spectral density.
\textcolor{black}{They are numerically calculated as $\Lambda(\omega_q) / \omega_q = -1.35\times10^{-2}$ ($-0.135$) and $\Lambda(-\omega_q)/\omega_q = -1.15\times10^{-2}$  ($-0.115$) for the weaker (stronger) coupling case, respectively.}

\subsection{\textcolor{black}{On the definition of non-Markovianity}}
In this section, we discuss the definition of the non-Markovianity.
The complete-positive-divisibility (CP-divisibility) of the mapping~\cite{Breuer2002} has been widely accepted for the definition of the Markovianity, and various measures have been proposed in a number of previous studies to detect the CP-divisibility.
We refer the readers to the review article~\cite{RivasRPP2014} for more details about the characterization of the non-Markovianity.
It was demonstrated in the paper~\cite{RivasRPP2014} that the time differential equation of the RDO in the Lindblad form is CP-divisible, and therefore the LE [Eq.~\eqref{eq:LE}] and ULE [Eq.~\eqref{eq:ULE}] in this paper is definitely Markovian process.
To investigate the Markovianity of the HEOM, we calculate the Breuer--Laine--Piilo (BLP) quantifier $\mathcal{N}_\mathrm{BLP}$~\cite{RivasRPP2014} of the HEOM.
Although it was pointed out that the process with $\mathcal{N}_\mathrm{BLP} = 0$ is not always CP-divisible~\cite{RivasRPP2014}, the process with $\mathcal{N}_\mathrm{BLP} > 0$ is strictly CP-indivisible and therefore is the non-Markovian process.
We found that the BLP quantifier for HEOM is positive, and we concluded that the exact dynamics without any approximations, which are obtained with the HEOM method, are the non-Markovian process in this study.
In the following, properties that are found in the HEOM results while not in the LE and ULE results are identified with the non-Markovian effects.
For more details of the calculation of $\mathcal{N}_\mathrm{BLP}$, see Appendix~\ref{sec:app}.

\subsection{Markovianity versus non-Markovianity}
\begin{figure}[h]
    \centering
    \includegraphics[width=\linewidth]{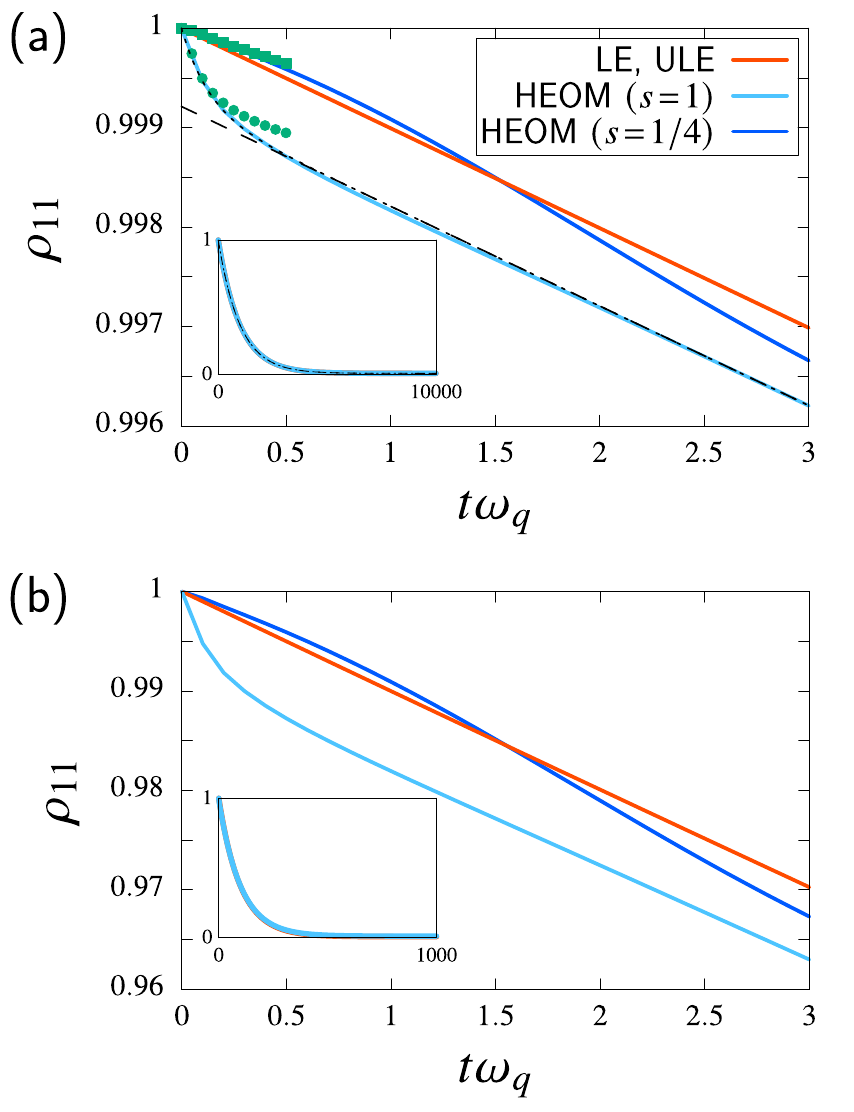}
    \caption{
        Dynamics of the population relaxation of the excited state $\rho_{11} (t)$ calculated with the LE, ULE and HEOM in the short-time region $0 \leq t \omega_q \leq 3$ (the inset exhibits the whole dynamics up to (a) $t \omega_q\leq 10000$ and (b) $t \omega_q \leq 1000$).
        \textcolor{black}{(a) The weaker ($2\pi\hbar\kappa = 10^{-3}$) and (b) the stronger ($2\pi\hbar\kappa = 10^{-2}$) coupling cases are depicted.}
        In the HEOM case, the Ohmic [$s=1$ in Eq.~\eqref{eq:SD}] and sub-Ohmic ($s=1/4$) spectral density are considered. The results of the LE and ULE coincide, and only a single curve (red curve) is shown.
        \textcolor{black}{The decay rate $\gamma_\mathrm{r}$ takes a same value irrespective of the exponent $s$, and therefore the curve does not change with $s$ in the LE and ULE cases.
        For this reason, the value of $s$ is omitted in the legend of LE and ULE.}
        The dashed and dotted black curves in (a) are the approximate curves of the Ohmic case with a single exponential function [Eq.~\eqref{eq:s_exp}] and a sum of two exponential functions [Eq.~\eqref{eq:d_exp}], respectively. The green circles and squares in (a) indicate the universal decoherence [Eq.~\eqref{eq:UD}] with the Ohmic and sub-Ohmic spectral density.
        \label{fig:population}
    }
\end{figure}

In Fig.~\ref{fig:population}, we depict the dynamics of the population relaxation of the excited state $\rho_{11}(t)$ that is numerically obtained with the LE, ULE and HEOM, respectively. For the FP-HEOM calculation, the Ohmic ($s=1$) and sub-Ohmic ($s=1/4$) spectral density are considered. We adopt $\hat{\rho}_S(t_0=0) = \ketbra{1}{1}$ as the initial states. As mentioned above, the dynamics of the LE and ULE coincide.
\textcolor{black}{In addition, the decay rate $\gamma_\mathrm{r}$ takes a same value irrespective of the exponent $s$.
The results do not change with $s$ in the LE and ULE cases, and therefore we do not explicitly mention the value of $s$ in the LE and ULE cases here.}

First, we focus on the analysis of the weaker coupling case, $2\pi\hbar\kappa = 10^{-3}$.
As illustrated in the inset of Fig.~\ref{fig:population}(a), the dynamics of the LE and FP-HEOM in the intermediate- to long-time region are qualitatively same. The same is true for the dependence on the exponent $s$ of the spectral density which causes qualitatively negligible effects. Quantitatively, the maximum difference of the population $\rho_{11}$ between the LE case and FP-HEOM case is on the order of $10^{-3}$.
We discuss the equilibrium state in detail in Sec.~\ref{sec:longTime}

\textcolor{black}{For the fault-tolerant quantum computation, according to common wisdom fidelities greater than $0.9999$ are required.
We infer from this constraint that even differences on the order of $10^{-3}$ between the results of LE/ULE and HEOM found in our study is significant.
We should not optimistically ignore this difference in order to aid the development of the practical quantum computer.}

In the short-time domain ($\omega_q t\lesssim 3$), the exact quantum dynamics cannot be described within the LE and ULE due to the underlying time coarse graining. In fact, both approaches predict a mono-exponential decay with rate $\gamma_{\rm r}$, while for very short times $\omega_q t<1$,
a fast decay of the population is observed in the Ohmic case. The comparison of corresponding numerical data with benchmark data (HEOM) in Fig.~\ref{fig:population} reveal that for the Ohmic case ($s=1$) and on timescales $\omega_q t\gtrsim 1$ a single exponential approximation according to
\begin{align}
    \bar{\rho}^{(1)}_{11}(t) = A^{(1)}\exp[-B^{(1)}t] + C^{(1)},
    \label{eq:s_exp}
\end{align}
does indeed capture the dynamics quite accurately. One finds parameters $A^{(1)} = 0.992$, $B^{(1)}/ \omega_q = 1.01\times10^{-3}$, $C^{(1)} = 7.15\times10^{-3}$, in line with the  predicted relaxation dynamics  $\gamma_\mathrm{r} / \omega_q = 1.01\times10^{-3}$. In the short-time domain, an approximation including two exponentials, i.e.,\ 
\begin{align}
    \bar{\rho}^{(2)}_{11}(t) = A^{(2)}_1\exp[-B^{(2)}_1t]
    + A^{(2)}_2\exp[-B^{(2)}_2t] + C^{(2)}\, ,
    \label{eq:d_exp}
\end{align}
provides a sufficiently precise description with $A^{(2)}_1 = 0.992$, $B^{(2)}_1 /\omega_q= 1.01\times10^{-3}$ and $A^{(2)}_2 = 7.90\times10^{-4}$, $B^{(2)}_2 /\omega_q= 7.94$, and $C^{(2)} = 7.15\times10^{-3}$. Before we discuss this in more detail, we turn to the sub-Ohmic case.

There, the population dynamics shows a nonmonotonous behavior.
It exceeds LE/ULE predictions until times $\omega_q t \simeq 1.5$, while it becomes smaller beyond. Subsequent oscillatory behavior around the LE/ULE data is observed for even longer times (not shown) but the absolute quantitative difference gradually decreases towards very long times. This oscillatory behavior clearly displays the {\em limitation of the Born--Markov approximation} and is due to time retarded feedback in the qubit--reservoir interaction. 

Now, coming back to the short-time region, for $t \omega_q \ll 1$ the contribution of the system dynamics to the total dynamics is negligible. This assumption implies $\hat{H}_S \simeq 0$ so that the dynamics of the population is governed by the coupling to the reservoir only
\begin{align}
    & \rho_{11}(t) \\
    = & \frac{1}{2} \left(1 + \exp \left[-4 \int_{t_0}^{t} dt' \int_{t_0}^{t'} dt''
    \mathrm{Re}\{C(t'-t'')\}\right] \right). \\
    \label{eq:UD}
\end{align}
This behavior is referred to as ``universal decoherence''~\cite{TuorilaPRR2019,BraunPRL2001}. 
Results are displayed for both Ohmic and sub-Ohmic reservoirs in Fig.~\ref{fig:population}(a) as the green circles and squares, respectively.
We found that indeed the sharp drop of population in this time domain is very well captured up to the time $t\omega_q \leq 0.1$ in the Ohmic case. The short-time behavior of the sub-Ohmic case is also well described with the universal decoherence. Because the impact of the system Hamiltonian is approximately absent in this time region, a perturbative approach with respect to $\hat{V}$ cannot be applied. This indicates another break-down of the Born--Markov approximation.

We thus conclude that both the fast initial decay and the oscillatory behavior of the population relaxation are  signatures of the non-Markovianity as they can neither be described by the LE (Markov approximation+RWA) and the ULE (Markov approximation only). 

\textcolor{black}{Next, we investigate the stronger coupling case ($2\pi\hbar\kappa = 10^{-2}$) in Fig~\ref{fig:population}(b).
The profile of Fig.~\ref{fig:population}(b) is qualitatively same as Fig.~\ref{fig:population}(a), except that the scale of the vertical axis is ten times greater in Fig.~\ref{fig:population}(b) than in Fig.~\ref{fig:population}(a).
This indicates that the maximum difference is on the order of $10^{-2}$, which is more significant than the weaker coupling case:
the violation of the Born--Markov approximation is more significant when the coupling strength is increased, and it is preferable to use HEOM approach when the coupling strength is not sufficiently small.}

\textcolor{black}{The inset of Fig.~\ref{fig:population}(b) shows that the population decay is approximately ten times faster than the weaker coupling case.
This is due to the ten times stronger system--bath coupling in Fig.~\ref{fig:population}(b) than in Fig.~\ref{fig:population}(a).}
\begin{figure}[h]
    \centering
    \includegraphics[width=\linewidth]{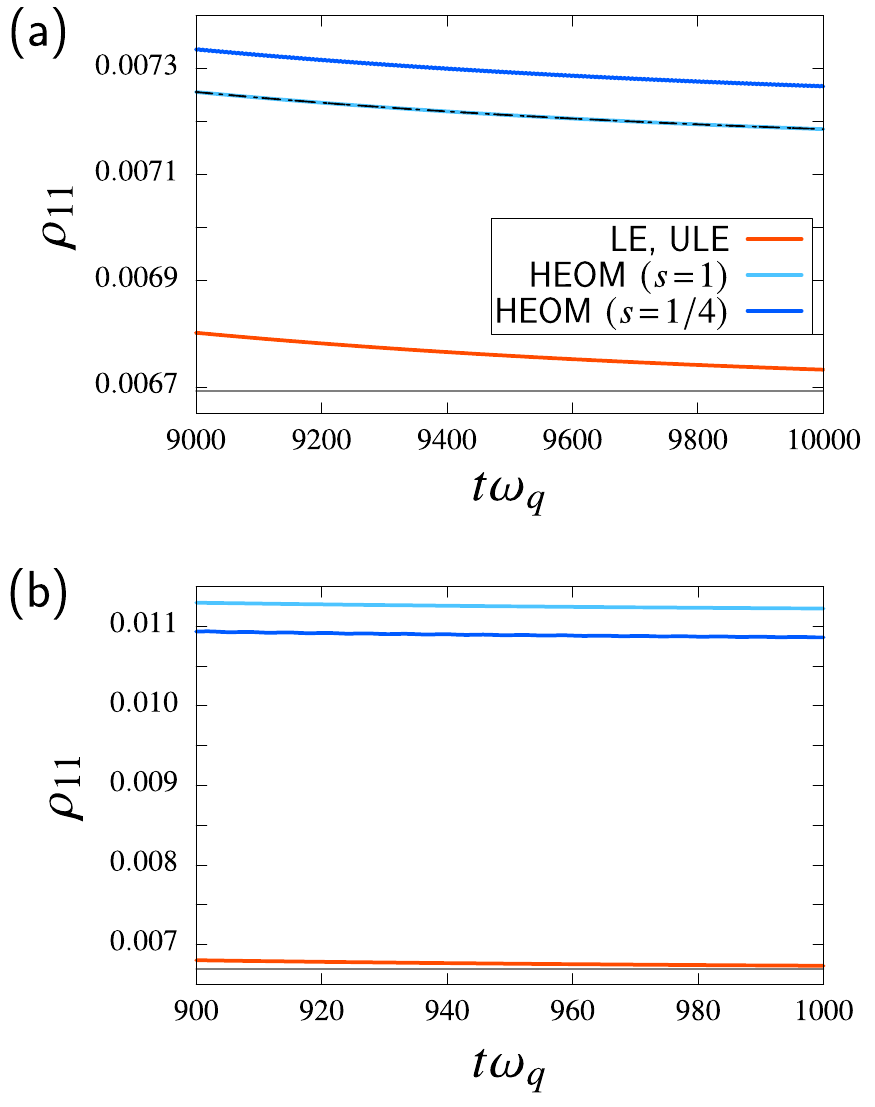}
    \caption{Dynamics of the excited-state population $\rho_{11}(t)$ in the long-time region (a) $9000 \leq t \omega_q \leq 10000$ and (b) $900 \leq t \omega_q \leq 1000$. The same curves as Fig.~\ref{fig:population} are depicted, and gray horizontal line is added to indicate the Boltzmann distribution of the bare system, $\rho^{eq,\mathrm{LE}}_{11}$.
    (a) The weaker ($2\pi\hbar\kappa = 10^{-3}$) and (b) the stronger ($2\pi\hbar\kappa = 10^{-2}$) coupling case are depicted.}
    \label{fig:populationEq}
\end{figure}
\subsection{\textcolor{black}{Long-time behavior of the density matrices}}\label{sec:longTime}
Here, we discuss the long-time behavior of the RDO.
Because we adopt the time-evolution equations in this study, we cannot obtain the equilibrium states strictly:
the difference of the RDO with respect to time is not zero at any time due to the computation.
For the HEOM, steady states are numerically obtained by solving the equation $\partial \hat{\rho}_{\vec{m}, \vec{n}}(t) / \partial t = 0$ in a self-consistent manner~\cite{zhang17sciheom}, but this method is computationally expensive.
For these reasons, we consider the states at a long time whose difference with respect to time is negligibly small as the equilibrium states.
Figure~\ref{fig:populationEq} displays the dynamics of Fig.~\ref{fig:population} in the long-time region.
In Fig.~\ref{fig:populationEq}(a), the absolute difference of the population $|\rho_{11}(t\omega_q=10000) - \rho_{11}(t\omega_q=9999)|$ is less than $10^{-6}$, and we consider the system at $t\omega_q = 10000$ as the equilibrium states.
The same discussion is applied to the stronger coupling case at the time $t\omega_q = 1000$, and we consider the system at this time as the equilibrium states in this case.

The gray line in Fig.~\ref{fig:populationEq} is the Boltzmann distribution of the bare system, $\rho^{eq, \mathrm{LE}}_{11}$.
As discussed above, the equilibrium distribution obtained with the LE and ULE is analytically the Boltzmann distribution with respect to the bare system Hamiltonian.
The small difference of the population between the Boltzmann distribution and the LE/ULE result at $t\omega_q = 10000$ in Fig.~\ref{fig:populationEq}(a) and $t\omega_q=1000$ in Fig.~\ref{fig:populationEq}(b) (approximately $4\times 10^{-5}$) implies the validity of our adoption of the equilibrium states.

The exact equilibrium state obtained with the FP-HEOM approach originates from the total Hamiltonian as $\hat{\rho}^{eq, \mathrm{exact}} = \mathrm{tr}_B\{e^{-\beta \hat{H}_\mathrm{tot}}\}/Z$ ($Z = \mathrm{tr}\{e^{-\beta\hat{H}_\mathrm{tot}}\}$), which is in principle different from the Boltzmann distribution of the bare system.
The difference of the equilibrium population between the LE/ULE result and HEOM one is on the order of $10^{-4}$ in Fig.~\ref{fig:populationEq}(a).
This relatively small difference between $\hat{\rho}^{eq, \mathrm{LE}}$ and $\hat{\rho}^{eq, \mathrm{exact}}$ is originated from the small coupling strength between the system and heat bath, $2 \pi \hbar \kappa = 10^{-3}$.
For the larger coupling case in Fig.~\ref{fig:populationEq}(b), $2\pi\hbar\kappa = 10^{-2}$, the enhanced difference on the order of $10^{-3}$ was observed.
For even larger coupling strengths, the difference is more significant~\cite{TuorilaPRR2019}, and we must seriously take into account the effects of $\hat{H}_I$.

It is interesting that the order of the population changes when the coupling strength changes:
in Fig.~\ref{fig:populationEq}(a), the excited-state population in the sub-Ohmic case is larger than in the Ohmic case, while that in the sub-Ohmic case is smaller than in the Ohmic case in Fig.~\ref{fig:populationEq}(b).

It is worth noting that the HEOM method is stable for the long-time simulations.
For example, algebraic decay of the two-point correlator of the spin-boson model, $S_{zz}(t) = \ev{\hat{\sigma}_z(t)\hat{\sigma}_z(0) + \hat{\sigma}_z(0)\hat{\sigma}_z(t)}/2$, was simulated with high accuracy with the FP-HEOM method in a previous study~\cite{XuPRL2022}.
\begin{figure}[h]
    \centering
    \includegraphics[width=\linewidth]{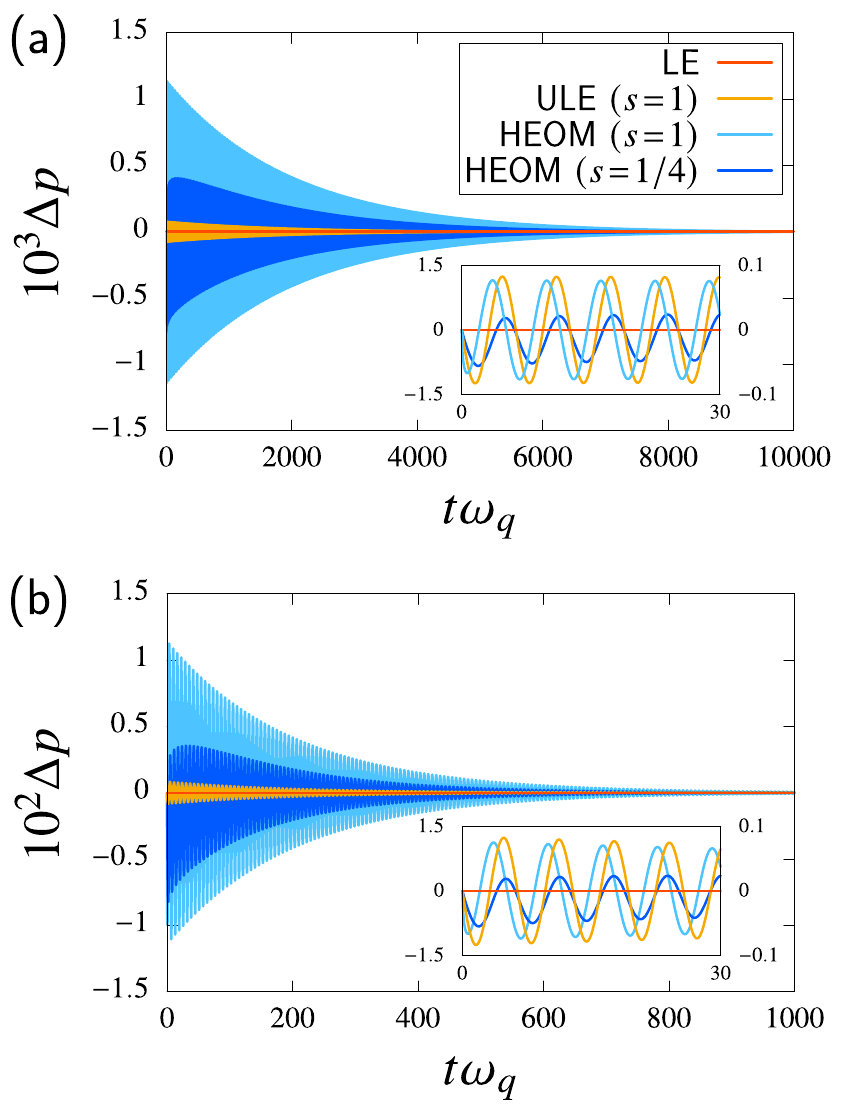}
    \caption{Difference, $\Delta p(t)$, of the ground-state population obtained after the application of the different pulse sequences, $\hat{X}_{-\pi/2}\hat{f}_t\hat{X}_{\pi/2}$ and $\hat{Y}_{\pi/2}\hat{f}_t\hat{Y}_{-\pi/2}$.
    \textcolor{black}{(a) The weaker ($2\pi\hbar\kappa = 10^{-3}$) and (b) the stronger ($2\pi\hbar\kappa = 10^{-2}$) coupling cases calculated with the LE, ULE and HEOM are depicted.}
    In the HEOM case, the Ohmic ($s=1$) and sub-Ohmic ($s=1/4$) spectral density are considered.
    \textcolor{black}{As a representative, the Ohmic case is depicted for the ULE case.
    The value $\Delta p(t)$ is always zero in the LE case irrespective of the exponent $s$, and we omit the value of $s$ in the legend of LE case.}
    (Inset) Time evolution of $\Delta p(t)$ in the short-time region. In the ULE case, the amplitude is scaled up to depict the behavior clearly. The corresponding scale is given by the second axis.
    \label{fig:difference}}
\end{figure}

\subsection{Experimental protocol: RWA versus non-Markovianity}
Next, we consider an experimental method proposed in a previous study to probe differences between  Markovianity and non-Markovianity in open qubit dynamics~\cite{GulasciPRB2022}. In that study, Ramsey experiments with two different pulse sequences are explored: One sequence consists of  pulses that rotates the qubit about the $x$ axis, while the other one consists of pulses that rotate the qubit about the $y$ axis. These pulse sequences are expressed as $\hat{X}_{-\pi/2}\hat{f}_t\hat{X}_{\pi/2}$ and $\hat{Y}_{\pi/2}\hat{f}_t\hat{Y}_{-\pi/2}$, respectively.
Here, $\hat{X}_{\theta}$ ($\hat{Y}_{\theta}$) corresponds to the ideal rotation operator with the angle $\theta$ about the $x$ ($y$) axis, and $\hat{f}_t$ indicates the time evolution without the pulses but with the heat bath. Note in passing that the difference of the sign of the pulse sequences between this study and the previous one~\cite{GulasciPRB2022} is based on a different sign in the Hamiltonian $\hat{H}_S$ in Eq.~\eqref{eq:TLS}. 

Assuming that the initial states of the systems before the pulse application are given by $\ketbra{0}{0}$, we evaluate the population of the ground state after the pulse sequences as follows: For the sequence $\hat{X}_{-\pi/2}\hat{f}_t\hat{X}_{\pi/2}$, we have
\begin{align}
    \varrho_{00}^{(X)}(t) = \frac{1}{2}
    + \mathrm{Im}\{\mel{0}{\rho_S^y(t)}{1}\},
    \label{eq:rhoX}
\end{align}
and for the sequence $\hat{Y}_{\pi/2}\hat{f}_t\hat{Y}_{-\pi/2}$,
\begin{align}
    \varrho_{00}^{(Y)}(t) = \frac{1}{2}
    + \mathrm{Re}\{\mel{0}{\rho_S^x(t)}{1}\}.
    \label{eq:rhoY}
\end{align}
Predictions for the RDO $\rho_S^x(t)$ and $\rho_S^y(t)$ are obtained within the LE [Eq.~\eqref{eq:LE_TLS}], the ULE [Eq.~\eqref{eq:ULE_TLS}], and the FP-HEOM [Eq.~\eqref{eq:HEOM}] with initial states $\hat{\rho}^x_S(t_0=0) = (\hat{1} + \hat{\sigma}_x)/2$ and $\hat{\rho}^y_S(t_0=0) = (\hat{1} + \hat{\sigma}_y)/2$, respectively. 
To quantify the potential difference between both sequences, it is convenient to introduce the difference of the populations as
\begin{align}
    \Delta p(t) & = \varrho_{00}^{(X)}(t) - \varrho_{00}^{(Y)}(t)
    \nonumber \\
    & = \mathrm{Im}\{\mel{0}{\rho_S^y(t)}{1}\}
    - \mathrm{Re}\{\mel{0}{\rho_S^x(t)}{1}\}\, .
\end{align}

Figure~\ref{fig:difference} displays the various time traces for $\Delta p(t)$.
Again, we first analyze the weaker coupling case [$2\pi\hbar\kappa = 10^{-3}$, Fig.~\ref{fig:difference}(a)].
In the HEOM calculation, both Ohmic and sub-Ohmic spectral density are considered. 
\textcolor{black}{In the LE and ULE cases, the Lamb shift is calculated on the basis of the Ohmic case ($s = 1$), as mentioned above.
However, in the LE case, the following argument holds for any value of the exponent $s$, and we do not explicitly mention the value of $s$ in the LE case.}
While the difference $\Delta p(t)$ in the LE case, derived from Eq.~\eqref{eq:LE_TLS}, is always zero, nonzero values are observed in the ULE and HEOM cases. Remarkably, the amplitudes of $\Delta p(t)$ are substantially larger for the benchmark data (HEOM) compared to the ULE case: the maximum absolute value is approximately $1 \times 10^{-3}$ for the HEOM with the Ohmic spectral density compared to $8 \times 10^{-5}$ for the ULE.

In the sub-Ohmic case, FP-HEOM data predict somewhat smaller amplitudes (approximately $8 \times 10^{-4}$) together with a characteristic asymmetric behavior with respect to $\Delta p = 0$, see  inset of Fig.~\ref{fig:difference}(a): the mean value of local maxima and minima is negative. In the Ohmic case, a weaker asymmetric behavior was found only in the short-time region, $t \omega_q \lesssim 1$. There, the amplitude of the first local minimum is slightly smaller than the amplitude of the subsequent oscillation. Our results indicate that an asymmetric behavior lasts for a longer time when the relative portion of low frequency modes increases (smaller $s$). Similar asymmetric behavior was found in a previous study~\cite{GulasciPRB2022}.

The frequency of oscillations in $\Delta p(t)$ carries also valuable information about the relevant qubit timescale in presence of reservoirs. From the inset of Fig.~\ref{fig:difference}(a), we retrieved the following frequencies: $\omega_{\Delta p} / \omega_q= 0.998$ for ULE, $0.999$ for FP-HEOM with the Ohmic spectral density and $1.00$ for sub-Ohmic spectral density, respectively. Note that since the initial phase of the oscillations is different in these three cases, local maxima and minima are observed at different times. We can as well extract the qubit's effective Larmor frequency from the Fourier transform of $\varrho^{(X)}_{00}$ and $\varrho^{(Y)}_{00}$ which are indeed in agreement with the respective frequencies $\omega_{\Delta p} $.

To gain further insight into the small deviations of these frequencies to the bare qubit transition frequency $\omega_q$, we consider the Lamb shift induced by the quantum reservoir. For the cases of LE and ULE, the predictions for modified qubit frequencies  $\tilde{\omega}$ are given through the function $\Lambda$ via $\tilde{\omega}-\omega_q=\Lambda(\omega_q)-\Lambda(- \omega_q)$; explicit results coincide indeed with the values for $\omega_{\Delta p}$.
\textcolor{black}{Note that the effective Larmor frequency is expressed as $\sqrt{\tilde{\omega}^2-\Delta^2}$ in the ULE case, but $\Delta  / \omega_q \lesssim 10^{-4}$ is small in our case ($2\pi\hbar\kappa = 10^{-3}$), and the main contribution of the frequency shift in the ULE is the Lamb shift.}

Predictions of the Lamb shift within the FP-HEOM for 
weakly coupled Ohmic reservoirs ($\hbar\kappa \ll 1$) are very well described within the framework of the noninteracting-blip approximation (NIBA)~\cite{Weiss2012,TuorilaPRR2019}. Then, for a spectral density of the form $J(\omega) = \kappa \omega e^{-|\omega|/\omega_c}$, one derives
\begin{align}
    \tilde{\omega}^2 = \omega_\mathrm{eff}^2
    \left\{1 \!+\! 2\mathcal{K} \left[\mathrm{Re}\left\{\psi\!
    \left(\frac{i\beta \hbar \omega_\mathrm{eff}}{2\pi}\right)\!\right\}
    - \ln\!\left(\frac{\beta\hbar\omega_\mathrm{eff}}{2\pi}\right)
    \right]\right\}. \\
    \label{eq:LS_NIBA}
\end{align}
Here, $\mathcal{K} = 2 \hbar \kappa$ is the Kondo parameter, and $\omega_\mathrm{eff}$ is given by
\begin{align}
    \omega_\mathrm{eff} = [\bar{\Gamma}(1-2\mathcal{K})
    \cos(\pi \mathcal{K})]^{1/[2(1-\mathcal{K})]}
    (\omega_q/\omega_c)^{\mathcal{K}/(1-\mathcal{K})}\omega_q.
\end{align}
The functions $\psi(x)$ and $\bar{\Gamma}(x)$ are the digamma and gamma functions, respectively. Using the chosen parameter values, the effective Larmor frequency is given by $\tilde{\omega}/\omega_q = 0.999$ in full agreement with $\omega_{\Delta p} / \omega_q$ in the Ohmic case, although the form of the spectral density is different. It was reported previously~\cite{TuorilaPRR2019} that the Lamb shift can be evaluated precisely with Eq.~\eqref{eq:LS_NIBA} irrespective of the cutoff function in the weak coupling regime.

These findings verify that  $\omega_{\Delta p}$ is determined by the effective Larmor frequency $\tilde{\omega}$ and that for weak coupling already a second order estimate provides a quite good description. Interestingly, the frequency difference between ULE, Ohmic, and sub-Ohmic cases is relatively small given that the amplitude differs by about an order of magnitude. If the RWA is not performed, one can obtain from the ULE precise information about the frequency of $\Delta p(t)$ (not about the amplitude though). Thus, the RWA plays a significant role for the estimation of $\Delta p(t)$.

This brings us back to the previous study~\cite{GulasciPRB2022}, where results with the LE and a TCL master equation, which can also describe non-Markovian dynamics (while the Born approximation is still imposed), were studied at zero temperature. It was reported that the difference $\Delta p(t)$ vanishes in the LE case and does not so in the case of the TCL master equation, in line with our results. It was then concluded that this difference clearly originates from non-Markovianity.

However, as we have seen above, this statement is questionable at nonzero temperatures: The difference $\Delta p(t)$ does {\em not} vanish for the ULE which imposes a Markov approximation but not the RWA. 
 In fact, at nonzero temperatures one can distinguish  Markovianity and RWA based on the LE and ULE, but this distinction becomes subtle at the zero temperature. Namely, then the ULE reaches the LE since $S_\beta(-\omega_q)=0$. We conclude that non-Markovianity cannot be identified only based on the observation of a finite $\Delta p(t)$. Other parameters such as temperature must be taken into account as well.

\textcolor{black}{Here, we study the behavior of $\Delta p(t)$ in the stronger coupling case, $2\pi\hbar\kappa = 10^{-2}$, in Fig.~\ref{fig:difference}(b).
Similar to Fig.~\ref{fig:population}, the profiles of Figs.~\ref{fig:difference}(a) and \ref{fig:difference}(b) are qualitatively same, except for the scale of axes:
the amplitude in the stronger case is ten times greater than in the weaker case, and the decay is ten times faster in case (b).
Note that the frequencies $\omega_{\Delta p}$ are almost the same for cases (a) and (b).
We emphasize again that when the coupling strength is not sufficiently small, the Born--Markov approximation is violated, and the HEOM approach provides more reliable results.}
\begin{figure}
    \centering
    \includegraphics[width=\linewidth]{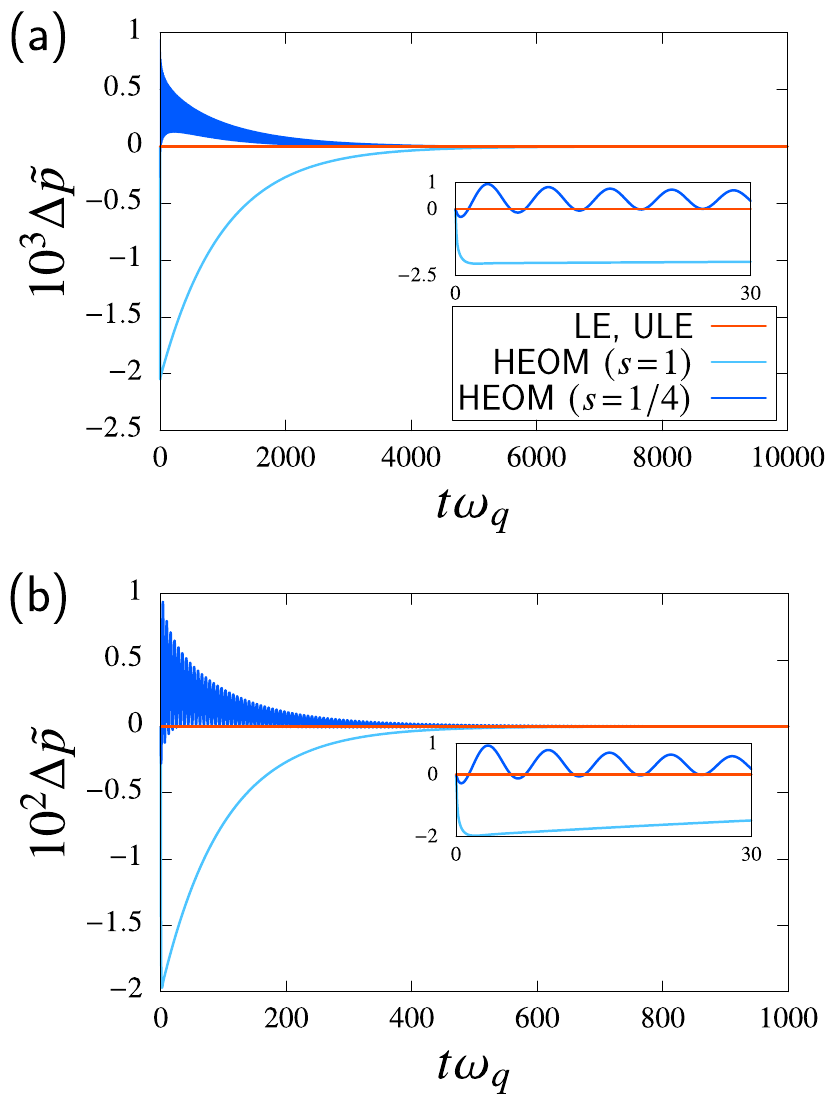}
    \caption{Difference, $\Delta \tilde{p} (t)$, of the ground-state population obtained after the different pulse sequences, $\hat{f}_t \hat{X}_\pi$ and $\hat{f}_t \hat{Y}_\pi$.
    (a) The weaker ($2\pi\hbar\kappa = 10^{-3}$) and (b) the stronger ($2\pi\hbar\kappa = 10^{-2}$) coupling cases calculated with the LE, ULE and HEOM are depicted.
    In the HEOM case, the Ohmic ($s=1$) and sub-Ohmic ($s=1/4$) spectral density are considered.
    The results obtained with LE and ULE coincide, and therefore only one line is shown.
    Also, this result is independent of the exponent $s$, and the value of $s$ is omitted in the legend of LE and ULE.
    (Inset) Time evolution of $\Delta \tilde{p}(t)$ in the short-time region.}
    \label{fig:differenceNew}
\end{figure}
\subsection{\textcolor{black}{Proposal of an experimental protocol for the distinction between the Markovianity and non-Markovianity}}
In the preceding section, we found that the previously proposed experimental method is not sufficient to distinguish the Markovianity from the non-Markovianity: Rather, it seems to be helpful to measure the impact of the RWA.
Here, we propose a new experimental method to detect the non-Markovianity:
a method to obtain a quantity that is zero in the LE and ULE cases, while nonzero in the HEOM cases.
Similar to the previous method, we consider pulse applications with the different rotation axis.
However, we utilize the $\pi$ pulse here, $\hat{X}_\pi$ and $\hat{Y}_\pi$, and the initial state is the equilibrium state, which we discussed in Sec.~\ref{sec:longTime}.
We assume that we experimentally obtain the equilibrium state
\begin{align}
    \hat{\rho}^{eq} = \frac{e^{-\beta\hat{H}_\mathrm{tot}}}
    {\mathrm{tr}\{e^{-\beta\hat{H}_\mathrm{tot}}\}}
    \label{eq:eqExact}
\end{align}
after the relaxation process without any operations to the qubit.
As discussed in Sec.~\ref{sec:longTime}, the equilibrium density operator is approximated with
\begin{align}
    \hat{\rho}^{eq} \simeq \frac{e^{-\beta \hat{H}_S}}
    {\mathrm{tr}\{e^{-\beta \hat{H}_S}\}} \otimes
    \frac{e^{-\beta \hat{H}_B}}{\mathrm{tr}\{e^{-\beta\hat{H}_B}\}}
    \label{eq:eqApprox}
\end{align}
within the Born--Markov approximation.
We aim to detect these difference by means of the method proposed below.

With the equilibrium initial state, we apply the $\pi$ pulse and then simply monitor the population relaxation of the system.
We consider two different $\pi$ pulses, $\hat{X}_\pi$ and $\hat{Y}_\pi$, and the sequences are symbolically expressed as $\hat{f}_t \hat{X}_\pi$ and $\hat{f}_t \hat{Y}_\pi$.
Explicitly, the ground-state populations we experimentally obtain are described as follows;
for the $\hat{f}_t \hat{X}_\pi$ sequence,
\begin{align}
    \tilde{\varrho}^{(X)}_{00}(t) = \mel{0}
    {\mathrm{tr}_B\{e^{-i \hat{H}_\mathrm{tot} t / \hbar} \hat{X}_\pi
    \hat{\rho}^{eq} \hat{X}_\pi^\dagger
    e^{i \hat{H}_\mathrm{tot} t / \hbar}\}}{0}, \quad
    \label{eq:rhoXNew}
\end{align}
and for the $\hat{f}_t\hat{Y}_\pi$ sequence,
\begin{align}
    \tilde{\varrho}^{(Y)}_{00}(t) = \mel{0}
    {\mathrm{tr}_B\{e^{-i \hat{H}_\mathrm{tot} t / \hbar} \hat{Y}_\pi
    \hat{\rho}^{eq} \hat{Y}_\pi^\dagger
    e^{i \hat{H}_\mathrm{tot} t / \hbar}\}}{0}. \quad
    \label{eq:rhoYNew}
\end{align}
For the LE and ULE, $\hat{\rho}^{eq}$ in Eqs.~\eqref{eq:rhoXNew} and \eqref{eq:rhoYNew} is approximated with Eq.~\eqref{eq:eqApprox}, and the time evolution $e^{\mp i \hat{H}_\mathrm{tot} t/ \hbar}$ is replaced with Eqs.~\eqref{eq:LE_TLS} and \eqref{eq:ULE_TLS}.
Note that because $\hat{\rho}^{eq}$ coincides in the LE and ULE cases, $\tilde{\varrho}^{(X)}_{00}(t)$ and $\tilde{\varrho}^{(Y)}_{00}(t)$ are same in both cases.

In the HEOM calculation, the RDO and ADOs at the time $t^{eq} = 10000 / \omega_q$ and $1000 / \omega_q$ are used as $\hat{\rho}^{eq}$ respectively for the weaker ($2\pi\hbar\kappa = 10^{-3}$) and stronger ($2\pi\hbar\kappa = 10^{-2}$) coupling case (see Sec.~\ref{sec:longTime}).
The pulse applications $\hat{X}_\pi \hat{\rho}^{eq} \hat{X}_\pi^\dagger$ and $\hat{Y}_\pi \rho^{eq} \hat{Y}_\pi^\dagger$ correspond to the operation $\hat{X}_\pi \hat{\rho}_{\vec{m},\vec{n}}(t^{eq}) \hat{X}_\pi^\dagger$ and $\hat{Y}_\pi \hat{\rho}_{\vec{m},\vec{n}}(t^{eq}) \hat{Y}_\pi^\dagger$ for all the ADOs.
The time evolution is evaluated with the HEOM [Eq.~\eqref{eq:HEOM}].

Similar to the previous method, we define the difference of the ground-state population between the different pulse sequences as
\begin{align}
    \Delta \tilde{p}(t) = \tilde{\varrho}^{(X)}_{00}(t) - \tilde{\varrho}^{(Y)}_{00}(t).
\end{align}
Figure~\ref{fig:differenceNew} displays the dynamics of $\Delta \tilde{p}(t)$ obtained with the LE/ULE and HEOM. In the HEOM case, the Ohmic and sub-Ohmic spectral density are considered.
The population obtained with the LE and ULE is independent of the exponent $s$, which is same as Figs.~\ref{fig:population} and \ref{fig:populationEq}, and therefore we do not explicitly mention the value of $s$ in the LE and ULE cases here.
Because the equilibrium state is the product state in the LE/ULE case, the difference of the rotation axis does not affect the following time evolution.
This leads to the time-independent zero value of $\Delta\tilde{p}(t)$.
By contrast, due to the term $\hat{H}_I$ in the exact equilibrium state [Eq.~\eqref{eq:eqExact}], the difference of the rotation axis affects the following time evolution, which leads to the nonzero value of $\Delta\tilde{p}(t)$ in the HEOM cases.
For the Ohmic case, fast monotonic decrease and slow monotonic increase was observed, while in the sub-Ohmic case, oscillatory behavior was found (see the insets of Fig.~\ref{fig:differenceNew}).

On the basis of the value of $\Delta \tilde{p}(t)$, we can distinguish the results of LE/ULE and HEOM.
This implies that the non-Markovianity is detected with the aid of the quantity $\Delta\tilde{p}(t)$.

Finally, we compare the weaker and stronger coupling cases.
Similar to the above results, the profiles of Figs.~\ref{fig:differenceNew}(a) and \ref{fig:differenceNew}(b) are qualitatively same except for the scale of the axes.
Similar to $\Delta p (t)$, the frequency of the oscillation of the sub-Ohmic case in the short-time region hardly changes with the change of the coupling strength. 
The signature of the non-Markovianity is enhanced when the strength of the system-bath coupling becomes large.

\section{Concluding remarks} \label{sec:conclusion}
\textcolor{black}{In this paper, we focused on the distinction between the Born--Markov approximation and rotating wave approximation, which have been widely adopted hand in hand for the studies of open quantum dynamics.}
We reviewed three equations of motion that describe the dynamics of open quantum systems, the approximate LE, ULE and the exact HEOM, and investigated how the approximations imposed to obtain the LE and ULE cause numerical errors in the dissipative dynamics of qubits.

Starting from the Born--Markov approximation, one obtains the ULE only on the basis of the properties of the heat bath (system agnostic), while we need additional information about the system to impose the RWA and to obtain the standard LE. These two examples of Lindblad equations demonstrate that there are a number of methods (approximations) to obtain the Markov equations for the open quantum systems in the Lindblad form.

Comparing the dynamics of the population relaxation from the qubit's excited state obtained with the LE, ULE, and FP-HEOM, we explored the errors caused by using the LE and ULE. When the coupling strength between system and heat bath is sufficiently small, results obtained with  LE and ULE on moderate to long timescales are qualitatively in agreement with exact results and only minor deviations asymptotically. Signatures of the limitation of these approximate equations are found in the short-time region of the dynamics and for non-Ohmic reservoirs: The universal initial decay and the oscillatory behavior cannot be expressed with a single monoexponential decay as predicted by LE/ULE. The stronger the coupling strength becomes, the more significant are differences between the approximate and exact results~\cite{TuorilaPRR2019, Nakamura18PRA}. Also, when we consider experiments with more complicated pulse sequences, those differences may become prominent~\cite{tanimura2015} even when the system--bath coupling is weak. For quantum computing experiments quantitative predictions with very high accuracy are demanded. In this respect numerically rigorous methods without approximations must definitely be used while Lindblad equations may typically provide only a relatively rough picture.
\textcolor{black}{For the improvement of approximate but computationally less expensive schemes, the above discrepancy between the HEOM and Lindblad equations must be considered, especially in the context of the simulations of quantum computation.}

Through numerical calculations of the difference of the ground-state population, $\Delta p(t)$, obtained from the different pulse sequences, we distinguished the impact of the Born--Markov approximation and the RWA on the qubit's dynamics. A finite $\Delta p(t)$ is observed even when the Born--Markov approximation is imposed at finite temperatures, in contrast to the zero temperature case considered previously in Ref.~\onlinecite{GulasciPRB2022}.
\textcolor{black}{We found that the RWA plays a more crucial role in the dynamics of $\Delta p$ than the Born--Markov approximation.}
The conclusion is that one cannot determine whether a system coupled to a heat bath obeys  Markovian or non-Markovian dynamics merely based on observing finite values for $\Delta p(t)$. While we found that the amplitude of $\Delta p(t)$ is substantially larger in the non-Markovian case, the threshold that unambiguously distinguishes it from the Markovian cannot easily be identified. It depends as well on system specific properties.

\textcolor{black}{To overcome this problem, we proposed a new experimental protocol.
Similar to the previously proposed protocol, we utilize the different rotation axis of the pulse and calculate the difference of the ground-state population.
The main difference from the previous protocol is that we use the $\pi$ pulse instead of the $\pi/2$ pulse and that we consider the correlated equilibrium state in terms of the total system$+$bath Hamiltonian for the initial state.
The obtained value with this protocol, $\Delta \tilde{p}(t)$, is always zero for the LE and ULE cases, and nonzero for the HEOM cases.
This unambiguously distinguishes the non-Markovianity from the Markovianity.
We hope that this new protocol is utilized for deeper understandings of the open quantum dynamics in the future experimental works.}
\section*{Acknowledgement}
The authors would like to thank M.~Xu for fruitful discussion and numerical assistance. This work was supported by the BMBF through QSolid and the DFG through AN336/17-1 (FOR2724).

\appendix
\section{\textcolor{black}{Breuer--Laine--Piilo (BLP) quantifier for the HEOM}}\label{sec:app}
In this appendix, we discuss the Breuer--Laine--Piilo (BLP) quantifier of the HEOM.
The BLP quantifier is defined as~\cite{RivasRPP2014,BreuerRMP2016}
\begin{align}
    \mathcal{N}_\mathrm{BLP} = \max_{\hat{\rho}_1, \hat{\rho}_2}
    \int_{\sigma > 0} dt \sigma(\hat{\rho}_1, \hat{\rho}_2, t),
    \label{eq:BLP}
\end{align}
where
\begin{align}
    \sigma(\hat{\rho}_1, \hat{\rho}_2, t) =
    \frac{d}{dt} D[\hat{\rho}_1(t), \hat{\rho}_2(t)]
\end{align}
and $D[\hat{\rho}_1(t), \hat{\rho}_2(t)] = ||\hat{\rho}_1(t) - \hat{\rho}_2(t)||/2$ is the trace distance between a pair of the RDOs at the time $t$ with different initial states, $\hat{\rho}_1(t)$ and $\hat{\rho}_2(t)$.

Numerically, the integral in Eq.~\eqref{eq:BLP} is evaluated as follows:
\begin{align}
    \int_{\sigma>0} dt \sigma(\hat{\rho}_1, \hat{\rho}_2, t)
    = \sum_{n=1}^{N} \mathtt{max}\left(0,
    \delta D[\hat{\rho}_1(n\delta t), \hat{\rho}_2(n\delta t)]\right). \\
    \label{eq:BLPComput}
\end{align}
Here, the finite difference of the trace distance is defined as
\begin{align}
    \delta D[\hat{\rho}_1(n \delta t), \hat{\rho}_2(n\delta t)] = & 
    D[\hat{\rho}_1(n \delta t), \hat{\rho}_2(n \delta t)] \\
    & - D[\hat{\rho}_1((n-1) \delta t), \hat{\rho}_2((n-1) \delta t)],
\end{align}
and the step size and total step for the numerical integration are given by $\delta t$ and $N$, respectively.
In this study, $\delta t \omega_q$ is set to $0.1$.
The function $\mathtt{max}(a, b)$ returns the larger value of $a$ and $b$.

Although we need to consider all the pairs of RDOs to obtain the exact value of $\mathcal{N}_\mathrm{BLP}$, we can demonstrate that the process is non-Markovian only by obtaining the finite value of Eq.~\eqref{eq:BLPComput} for some pair of RDOs~\cite{RivasRPP2014,BreuerRMP2016}.
We calculated the value of Eq.~\eqref{eq:BLPComput} of the HEOM for the pair of $\hat{\rho}^y_S(t)$ and $\hat{\rho}^x_S(t)$ in Eqs.~\eqref{eq:rhoX} and \eqref{eq:rhoY}.
We set the total step for the integral as $N = 10000 / (\delta t \omega_q)$ for the weaker coupling case and $N = 1000 / (\delta t \omega_q)$ for the stronger coupling case, and obtained the value $0.207$ for the Ohmic bath and $3.37\times10^{-3}$ for the sub-Ohmic bath in the weaker coupling case, and $0.204$ for the Ohmic bath and $3.43\times10^{-3}$ for the sub-Ohmic bath in the stronger coupling case.
From these values, we concluded that the BLP quantifier $\mathcal{N}_\mathrm{BLP}$ for the HEOM is not zero and that the exact process without any approximations is the non-Markovian process in our case.
\bibliography{reference,qubit}
\end{document}